\documentclass{aa610}

\usepackage{amsmath}
\usepackage{natbib}
\usepackage{graphicx}
\usepackage{txfonts}
\usepackage{color}
\usepackage{deluxetable}
\usepackage{amssymb}

\newcommand{\M}{\bold{M}}
\newcommand{\Hv}{\bold{H}}
\newcommand{\D}{\bold{D}}
\newcommand{\V}{\bold{V}}
\newcommand{\C}{\bold{C}}

\newcommand{\om}{\Omega_M}
\newcommand{\1}{$x_{1}$}

\newcommand{\NumberOfNearbySNe}{52 }

\newcommand{\New}[1]{#1}

\begin{document}
\title{SALT2: using distant supernovae to improve the use of Type Ia 
supernovae as distance indicators
\thanks{
Based on observations obtained with MegaPrime/MegaCam, a joint project
of CFHT and CEA/DAPNIA, at the Canada-France-Hawaii Telescope (CFHT)
which is operated by the National Research Council (NRC) of Canada,
the Institut National des Sciences de l'Univers of the Centre National
de la Recherche Scientifique (CNRS) of France, and the University of
Hawaii. This work is based in part on data products produced at the
Canadian Astronomy Data Centre as part of the Canada-France-Hawaii
Telescope Legacy Survey, a collaborative project of NRC and CNRS.
Based on observations 
obtained at the European Southern Observatory using the Very Large Telescope
on the Cerro Paranal (ESO Large Programme 171.A-0486).
Based on observations (programs GN-2004A-Q-19, GS-2004A-Q-11, GN-2003B-Q-9, and
GS-2003B-Q-8) obtained at the Gemini Observatory, which is
operated by the Association of Universities for Research in Astronomy,
Inc., under a cooperative agreement with the NSF on behalf of the
Gemini partnership: the National Science Foundation (United States),
the Particle Physics and Astronomy Research Council (United Kingdom),
the National Research Council (Canada), CONICYT (Chile), the
Australian Research Council (Australia), CNPq (Brazil) and CONICET
(Argentina).
}% end of thanks 
}
\titlerunning{SALT2}
\author{
J.~Guy\inst{1},
P.~Astier\inst{1},
S.~Baumont\inst{1},
D.~Hardin\inst{1},
R.~Pain\inst{1},
N.~Regnault\inst{1},
S.~Basa\inst{2},
R.G.~Carlberg\inst{3},
A.~Conley\inst{3},
S.~Fabbro\inst{4},
D.~Fouchez\inst{5},
I.M.~Hook\inst{6},
D.A.~Howell\inst{3},
K.~Perrett\inst{3},
C.J.~Pritchet\inst{7},
J.~Rich\inst{8},
M.~Sullivan\inst{3},
P.~Antilogus\inst{1},
E.~Aubourg\inst{8},
G.~Bazin\inst{8},
J.~Bronder\inst{6},
M.~Filiol\inst{2},
N.~Palanque-Delabrouille\inst{8},
P.~Ripoche\inst{5},
V.~Ruhlmann-Kleider\inst{8}
}
\institute{
LPNHE, CNRS-IN2P3 and Universit\'{e}s Paris VI \& VII, 4 place Jussieu,
75252 Paris Cedex 05, France
\and LAM, CNRS, BP8, Traverse du Siphon, 13376 Marseille Cedex 12, France % 2-Basa
\and Department of Astronomy and Astrophysics, University of Toronto,
50 St. George Street, Toronto, ON M5S 3H8, Canada                         % 3-Carlberg
\and CENTRA-Centro M. de Astrofisica and Department of Physics, IST, Lisbon, 
Portugal                                                                  % 4-Fabbro
\and CPPM, CNRS-IN2P3 and Universit\'e Aix-Marseille II, Case 907, 
13288 Marseille Cedex 9, France                                           % 5-Fouchez
\and University of Oxford Astrophysics, Denys Wilkinson Building, Keble Road, % 6-Hook
Oxford OX1 3RH, UK
\and Department of Physics and Astronomy, University of Victoria,       % 7-Pritchet
PO Box 3055, Victoria, BC VSW 3P6, Canada
\and DSM/DAPNIA, CEA/Saclay, 91191 Gif-sur-Yvette Cedex, France  % 8-Rich
%\and APC, Coll\`ege de France, 11 place Marcellin Berthelot, 75005 Paris, France
%\and Department of Physics and Astronomy, University of Victoria, 
%PO Box 3055, Victoria, BC VSW 3P6, Canada
%\and Universit\'e de Savoie, 73000 Chambery, France
%\and LBNL, 1 Cyclotron Rd, Berkeley, CA 94720, USA
%\and ESO, Alonso de Cordova 3107, Vitacura, Casilla 19001, Santiago 19, Chile
%\and CRAL, 9 avenue Charles Andre, 69561 Saint Genis Laval cedex, France 
%\and California Institute of Technology, Pasadena, California, USA
%\and LUTH,UMR 8102, CNRS and Observatoire de Paris, F-92195 Meudon, France
%\and Department of Physics, Stockholm University, Sweden
%\and IoA, University of Cambridge, Madingley Road, Cambridge, CB3 0EZ, UK
%\and Department of Physics, University of California Berkeley, Berkeley, CA 94720, USA
} 

\authorrunning{J. Guy et al, SNLS Collaboration}                         
\offprints{guy\@@in2p3.fr}
\date{Received Month DD, YYYY; accepted Month DD, YYYY}

\abstract{}{We present an empirical model of Type Ia supernovae
spectro-photometric evolution with time.}{The model is built using a
large data set including light-curves and spectra of both nearby and
distant supernovae, the latter being observed by the SNLS
collaboration.  We derive the average spectral sequence of Type Ia
supernovae and their main variability components including a color
variation law.  The model allows us to measure distance moduli in the
spectral range $2500-8000~\AA$ with calculable uncertainties,
including those arising from variability of spectral features.}{Thanks
to the use of high-redshift SNe to model the rest-frame UV spectral
energy distribution, we are able to derive improved distance estimates
for SNe~Ia in the redshift range $0.8<z<1.1$. The model can also be
used to improve spectroscopic identification algorithms, and derive
photometric redshifts of distant Type Ia supernovae.}{}
\keywords{supernovae: general - cosmology: observations}

\maketitle

\section{Introduction}

The evolution of luminosity or angular distance with redshift is an
 essential observable to constrain the equation of state of dark
 energy, responsible for the acceleration of the expansion of the
 Universe. Type Ia supernovae (SNe~Ia) are, today, among the best
 distance indicators.  They can be observed up to high redshifts, and
 provide a precise estimator, with a typical dispersion of 7\% on
 distance, when not limited by the measurement uncertainties (see
 e.g.~\citealt{Astier06}, hereafter A06).

As the number of SNe~Ia in Hubble diagrams increases, systematic
uncertainties are becoming the main limitation to the accuracy of
measurements of cosmological parameters with SNe~Ia.  Among the
potentially serious systematic uncertainties, the dominant ones are a
possible evolution of the supernova population, the photometric
calibration, the modeling of the instrument response and the
uncertainties arising from SN~Ia large spectral features, including
their possible supernova to supernova variations.  In this paper, we
aim at addressing principally the latter although the proposed method
makes it also easy to account for the modeling of the instrument and
to propagate the model uncertainties.

Various approaches to distance estimation have been proposed, using
light-curve shape parameters ($\Delta m_{15}$ or a stretch factor, see
e.g.~\citealt{Phillips93, Riess95, Perlmutter97}) or color information
(\citealt{Wang03}, \citealt{Wang05}), or both (\citealt{Riess96b},
\citealt{Tripp98}, \citealt{Guy05}).  None of these methods really
address the problem of uncertainties due to the variability of the
large features of SNe~Ia spectra.  In this respect, most methods rely
on the spectral sequence provided by~\citet{Nugent02} (hereafter
N02). This is a serious concern since this could possibly result in
sizable (common) systematic effects in the distance measurements.  In
\citet{Guy05} (hereafter SALT), we have applied broadband corrections
to the spectral sequence N02 as a function of phase, wavelength and a
stretch factor so that the spectra integrated in response functions
match the observed light-curves. This provided us with a tool to fit
the observed light-curves without correcting the data points since the
$K$-corrections were naturally built into the model (see e.g. N02 for
a definition of $K$-corrections).  This approach has been quite
successful when applied to estimating SNe~Ia distances at high
redshift. However it did not address the problem of variability of
spectral features either.

In this paper, we use a similar framework. Whereas we only used
 multi-band light curves to train the model in SALT, here we include
 spectroscopic data to improve the model resolution in wavelength
 space, be fully independent of the spectral sequence N02, and more
 generally extract the maximum amount of information from the current
 data sets.  Modeling the supernova signal in spectroscopic space
 ensures that the $K$-corrections are treated in a consistent manner
 since there is a single model to address both light-curves and
 spectra.  It also permits a coherent propagation of errors, from the
 fit of light-curves to distance estimate.  The model is allowed to
 vary as a function of phase and wavelength with a small number of a
 priori unknown intrinsic parameters and a color variation law which
 is also adjusted during the training process. The main goal of this
 approach is to provide the best ``average'' spectral sequence and the
 principal components responsible for the diversity of SNe~Ia, so that
 the model can account for possible variations in SNe~Ia spectra at
 any given phase.

The flux normalization of each supernova is a free parameter of the
model. Hence, we do not need to know their distances to train the
model. This allows us to use both nearby SNe which are not in the
Hubble flow and high-redshift ones without any prior on cosmology.
Using high-redshift supernovae permits to model the rest-frame UV
emission which is invaluable to improve distance estimates of
supernovae found at redshifts larger than 0.8.

In Sect.~\ref{sec:model} we present the model implementation.  The
supernova data sets used for training the model are described in
Sect.~\ref{sec:dataset}.  Some technical aspects of the training
procedure are then given in Sect.~\ref{sec:training}, and in
Sect.~\ref{sec:results}, we present some qualitative aspects of the
resulting model.  In an effort to improve distance estimates for a
cosmology application to SNe~Ia surveys, we quantify the remaining
variability beyond the principal components extracted in
Sect.~\ref{sec:residuals}.  We show how distance estimates of SNLS
distant SNe are improved with this approach in
Sect.~\ref{sec:improving} and discuss several other possible use of
the model in Sect.~\ref{sec:applications}.

\section{The Type Ia supernova spectral sequence model}
\label{sec:model}

We aim at modeling the mean evolution of the spectral energy
distribution (SED) sequence of SNe~Ia and its variation with a few
dominant components, including a time independent variation with
color, whether it is intrinsic or due to extinction by dust in the
host galaxy (or both). The following functional form for the flux is
used
\begin{eqnarray}
F(SN,p,\lambda) =  x_0 & \times & \left[ M_0(p,\lambda) +  x_1 M_1(p,\lambda) + ... \right]  \nonumber \\
& \times & \exp \left[ c CL(\lambda) \right] \label{eq:model}
\end{eqnarray}
where $p$ is the rest-frame time since the date of maximum luminosity
in B-band (the phase), and $\lambda$ the wavelength in the rest-frame
of the SN.  $M_0(p,\lambda)$ is the average spectral sequence whereas
$M_k(p,\lambda)$, for $k>0$, are additional components that describe
the main variability of SNe~Ia. $CL(\lambda)$ represents the average
color correction law. As for SALT, the optical depth is expressed
using a color offset with respect to the average at the date maximum
luminosity in B-band, $c = (B-V)_{MAX} - \left< B-V \right>$.  This
parametrization models the part of the color variation that is
independent of phase, whereas the remaining color variation with phase
is accounted for by the linear components. $x_0$ is the normalization
of the SED sequence, and $x_k$ for $k>0$, are the intrinsic parameters of
this SN (such as a stretch factor).
 To summarize, whereas $(\M_k)$ and $CL$ are properties of the
global model, $(x_k)$ and $c$ are parameters of a given supernova and
hence differ from one SN to another.

Except for the color exponential term, Eq.~\ref{eq:model} is
equivalent to a principal component decomposition. However, a
principal component analysis cannot be used since this would require
having an homogeneous and dense set of observations for each SN,
namely one spectro-photometric spectrum every 4--5 days, which is not
presently available (note that current ongoing SN programs such as the
SNfactory, \citealt{SNFAldering}, \New{the Carnegie Supernova Program,
\citealt{HamuyCSP2006}, the CfA Supernova program~\footnote{CfA
Supernova Group:\\ 
{\it cfa-www.harvard.edu/oir/Research/supernova/index.html}} and the
LOTOSS project~\footnote{The Lick Observatory and Tenagra Observatory
Supernova Searches: {\it astro.berkeley.edu/~bait/lotoss.html}}},
should provide such data in the coming years).  So we resorted to
using a method able to deal with missing data. The method used is
described in the next section.

\subsection{Model implementation}
\label{sec:implementation}
The phase space that we want to model (wavelength range times phase
range) is not covered by the set of observations of any given
supernova. We typically have for each supernova a limited set of
light-curves points observed with different filters and, for some
supernovae, one or several spectra at different phases.  However, when
using an ensemble of SNe, this phase space can be correctly sampled
and if the data set is large enough, several components can be
extracted.

In order to link the model defined by a limited set of parameters and
the SNe observations, we used a basis of functions, as function of
phase and wavelength $\left[f_i(p,\lambda)\right]$.  We used third
order B-splines (to ensure continuous second derivatives).  The actual
choice of the basis is irrelevant in the phase space regions which are
densely covered by data, as long as it provides a sufficient
resolution to follow the observed variability of the SED sequence as a
function of phase and wavelength.  Choosing another basis will modify
the model in regions where it is poorly constrained, such as very
early spectra ($p<-15$~days). As described in
section~\ref{sec:residuals}, those poorly constrained phase space
regions are identified after the training using a jack-knife
technique. In this framework, a model is a linear combination of the
basis functions and can be described by a vector $\M$.  Each
measurement at a given phase and wavelength $m(p_m,\lambda_m)$ is then
compared to the model with a vector $\Hv_m$ (with values
$H_{m,i}=f_i(p_m,\lambda_m)$), so that the expected value for the
model at $(p_m,\lambda_m)$ is the scalar product $\Hv_m^{\mathrm T}
\M$.

\subsection{The use of spectral information}
Most spectra of SNe~Ia available in the literature are not calibrated photometrically. 
Their flux calibration have broad-band systematic uncertainties.

One way to circumvent this difficulty consists in photometrically "re-calibrating" 
a given spectrum using the available light-curves for this particular SN.
However since the full SED sequence model is needed to derive an accurate interpolation between light 
curve points at the date of the spectroscopic observation, and the spectra are needed 
as well to accurately model the spectral features of SNe, 
the photometric "re-calibration" of the spectra has to be included in the global 
minimization procedure.
We have chosen to parameterize the "re-calibration" function with the exponential 
of a polynomial (to force positive corrected fluxes), with
the degree of the polynomial limited by the number of light curves for the SN, and the wavelength range of the spectrum. \New{This re-calibration function is applied to the model, for which the SN parameters are mostly constrained by the light-curves. Thanks to the simultaneous use of a large amount of SNe data, we do not need to have photometric observations at the same epoch as spectroscopic ones.}

Statistical errors are rarely provided with the spectra. We have evaluated 
them using the fact that all SNe spectral features are broadened due to the kinematics 
of the ejected matter. So we expect an intrinsic correlation length greater 
than $30 \AA$ (for a velocity range larger than about 2000 km.s$^{-1}$) which permits 
one to evaluate the photon noise in spectra (assumed to be white noise). 
Nonetheless, we scaled errors so that the weight 
of spectra was of order of that of light curves for which we expect 
lower systematic errors. This weighting, along with the resolution of the 
re-calibration function, is a bit arbitrary but can not be avoided at this 
stage due to the quality of currently available data sets.  

\section{The training supernova data sets}
\label{sec:dataset}
In this section we describe the data sets used for training the model.

In the proposed model (Eq.~\ref{eq:model}), the overall SED sequence normalization of 
 a given SN is a free parameter ($x_0$).
As a consequence,  it is possible to use both the very nearby supernovae data that are not in the 
Hubble flow ($z<0.001$) and the high redshift ones without adopting values of cosmological parameters.

Nearby supernovae have
much higher signal to noise than their distant counterparts over a much
wider range of phases.
One important difficulty however in using nearby SNe is that they 
suffer from potentially 
large systematic errors in the ultra-violet (UV), since the atmospheric 
extinction is strong and difficult to model in this wavelength range.
Including SNe from a large redshift range helps to sample homogeneously the rest-frame 
visible wavelength range with both photometric and spectroscopic data, especially 
in the rest-frame UV. Indeed, if only nearby SNe are used, photometric data  
do not cover the gaps between the central wavelength of the filter set 
used (mostly Johnson-Cousins UBVRI), and one relies only on spectra for interpolation 
between those bands, which may introduce (weak) systematic effects on $K$-corrections.

One may argue that possible evolution of the SNe~Ia with redshift might 
cause some problems with the modeling since objects at all redshifts are used to 
obtain the model. 
Actually, the model describes an average SN~Ia 
at an average redshift but evolution can still be studied.  
%it might even be useful in the sense that we can test evolution, 
For instance, without any a priori on the effect of evolution on SED sequence, one can look 
 at the variation of the $(x_k)$ parameters with redshift.

\subsection{The nearby supernova sample}
We use a sample of \NumberOfNearbySNe nearby supernovae (without restricting ourselves to 
very nearby ones as in SALT) listed table~\ref{tab:training_sample}. Those SNe were selected 
from the quality of their light-curve 
sampling where we basically require measurements before the date of maximum to ensure a good 
estimate of the luminosity at peak. 
A large fraction of those SNe light-curves come from~\citet{Hamuy96a,Riess96b} and
\citet{Jha05}. We did not consider 1991bg-like SNe~Ia (with very low stretches). 
They have such different light-curves and spectra that the linear model we consider cannot 
fit those along with other SNe~Ia. 
This is not a problem since we aim at modeling the bulk of the SNe~Ia population 
(and we do not expect to detect many of those objects at high redshift).

We do not use any spectra without photometric data for the same SN (at
least two light-curves in different filters), so that the date of
maximum, color and ($x_k$) can be determined. However, since spectra
are calibrated on the model and not on the photometric data, we do not
need simultaneous photometric observations; we just need enough
photometric observations to derive the SN parameters.  From the sample
of \NumberOfNearbySNe SNe, we were able to gather 264 spectra for 16
SNe. There are 10 spectral sequences (with more than 10 spectra),
namely 1989B, 1990N, 1991T, 1992A, 1994D, 1996X, 1998aq, 1998bu,
1999ee, and 2002bo (see Table~\ref{tab:training_sample} for the
complete list of spectra and their references).

All available UV spectra from the International Ultraviolet Explorer~\citep{IUE} were included. This is
very helpful since most high-redshift SNe spectra which cover the rest-frame UV range have a 
low signal to noise ratio.
For all spectra from ground-based observations, we do not consider any measurement below 3400~$\AA$ 
because of the strength 
of the atmospheric absorption in this spectral region.

\subsection{The high-redshift supernova sample}
\label{sec:highz_sample}
We used a set of 121 Type Ia supernovae light-curves obtained by the
Supernova Legacy Survey (SNLS) during the first 2 years of the survey
(see Table~\ref{tab:training_sample}). The light curves were obtained
with the same reduction pipeline as described in A06, but with new
images in the photometric fit, so that light-curves have more data
points and with improved statistical accuracy since the reference data
used to anchor the estimate of the galaxy brightness become deeper
with time (thanks to the rolling search observing strategy). All the
71 SNe used for cosmology in A06 were used, with 50 additional ones.

In addition to the light-curve points, we used 39 high-redshift SNLS
spectra obtained at VLT\New{~\citep{Basa07}} and
Gemini\New{~\citep{Howell05}} during the regular SNLS observation
programs, which aim at typing and measuring the redshifts of SN
candidates.  Obviously more spectra were recorded (at least one for
each SN) but we choose to use only those with negligible residual
contamination from the host galaxy.  The contribution from the host
galaxy was removed in the reduction procedure of the VLT
spectra~\citep{Baumont}.  For all spectra, the remaining contamination
was evaluated a posteriori using the model itself with the following
procedure: using the SN parameters retrieved from the fit of
light-curves, the model was fit on the SN spectrum with re-calibration
parameters, and an additional contribution of the host galaxy. We used
for this purpose templates of elliptical, S0, Sa, Sb and Sc galaxies,
the actual galaxy type was fitted at the same time as its
normalization.  All spectra with a non zero contribution of the galaxy
(at 68\% confidence level) were not used in the training sample.

\begin{figure}
\begin{center}
\begin{minipage}[c]{\linewidth}
\includegraphics[width=\linewidth]{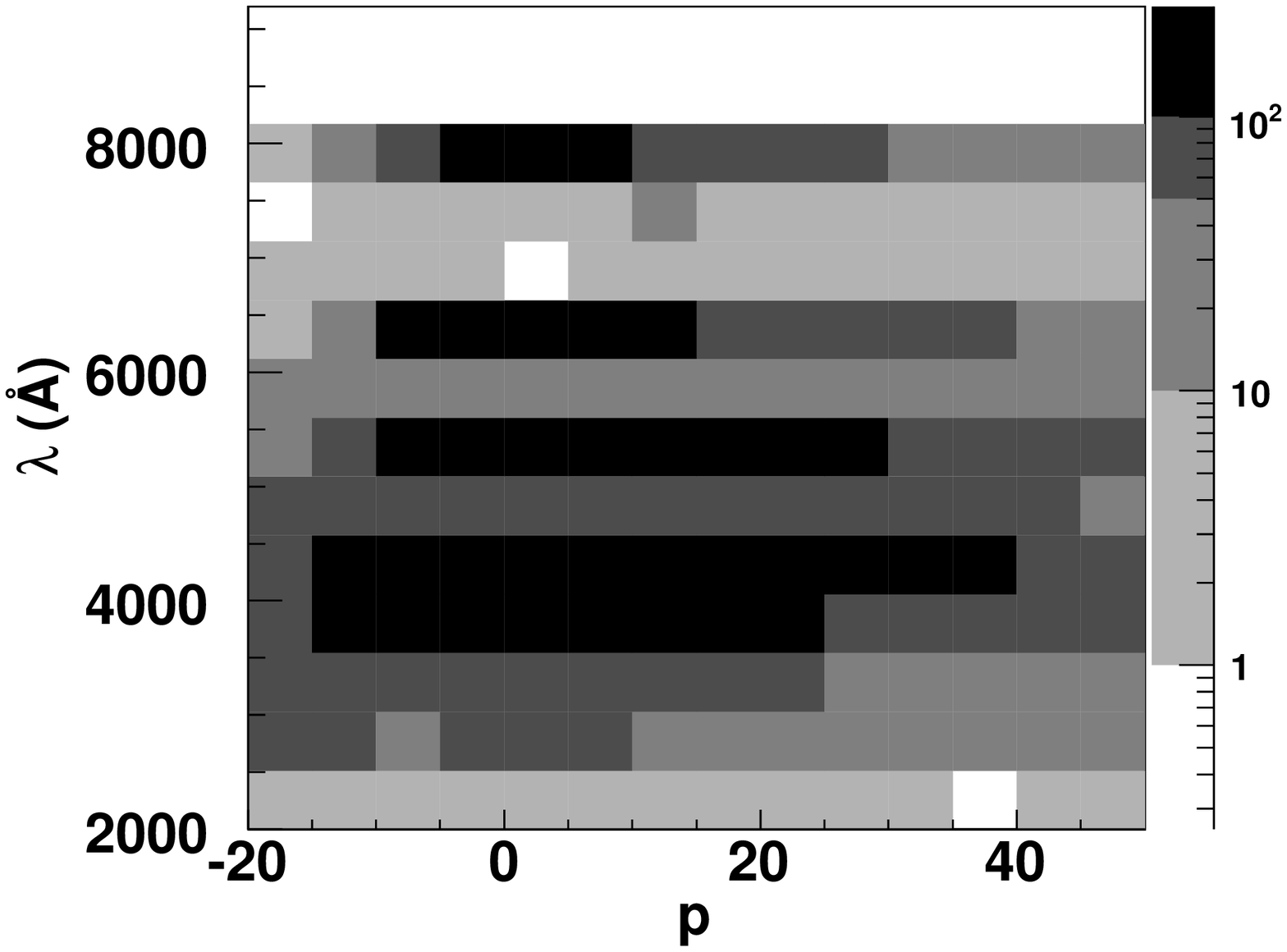}
\end{minipage}
\begin{minipage}[c]{\linewidth} 
\includegraphics[width=\linewidth]{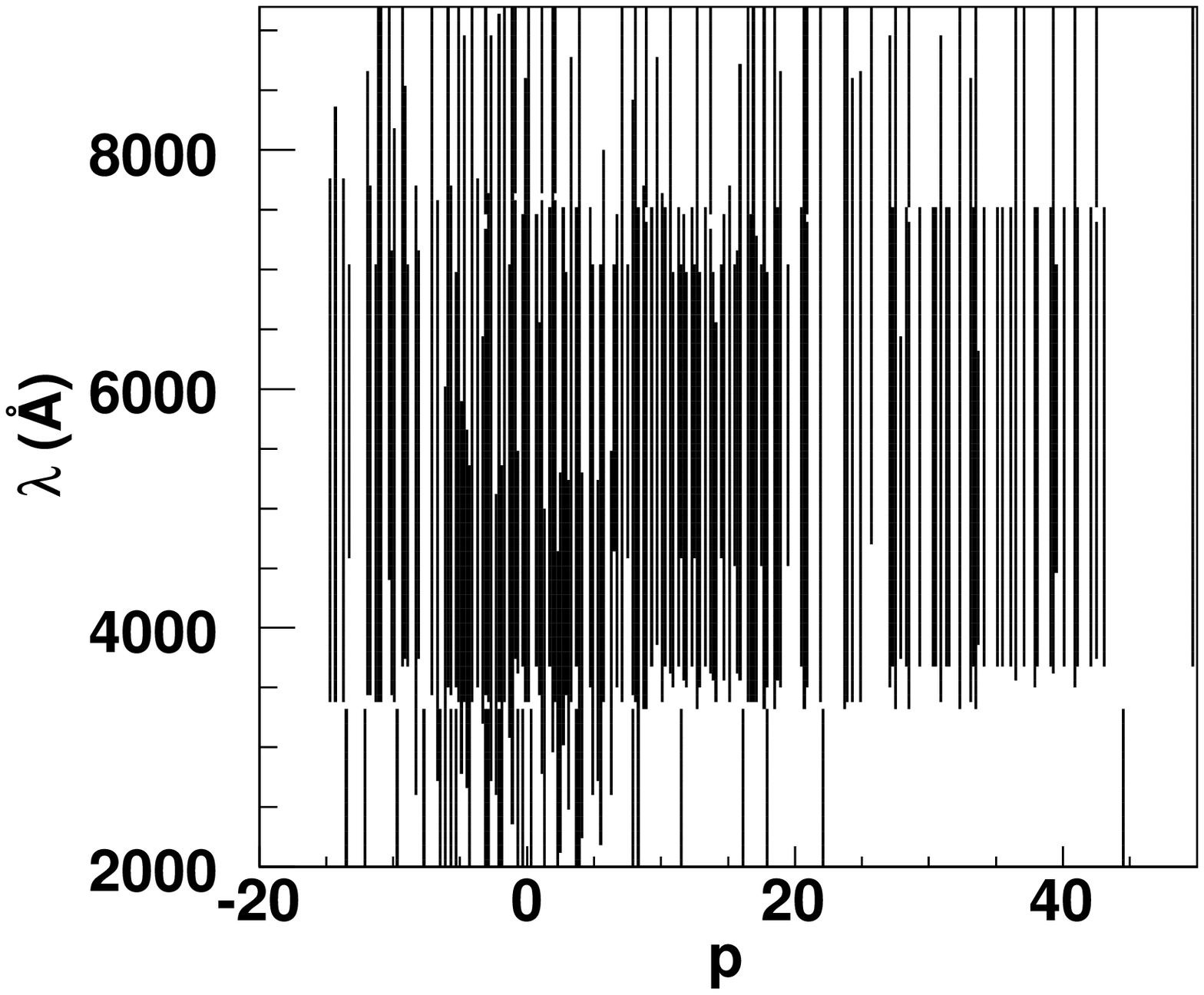}
\end{minipage}
\caption{Phase-space mapping by photometric data (top) and spectra
(bottom). For photometric observations, the rest-frame central
wavelength of the filter is considered.
\label{fig:coverage}}
\end{center}
\end{figure}

Figure~\ref{fig:coverage} shows the $(p,\lambda)$ phase-space region
covered by the photometric and spectroscopic data sets.  Since we do
not use infra-red photometric data, the re-calibration of spectra may
not be reliable for rest-frame wavelengths larger than 8000~$\AA$,
which is the central wavelength of the $I$-band filter. Also, we have
little spectroscopic information in the UV for phases earlier than
$-10$~days or greater than $10$~days since the spectroscopic
observations of the SNLS are designed to be as close as possible to
the date of maximum luminosity.  The few late UV spectra we have in
our sample come from IUE database~\citep{IUE}.
  
\section{Training the model}
\label{sec:training}

\subsection{The training procedure}

The convergence process consists in minimizing a $\chi^2$ that permits
 the comparison of the full data set with the model of
 equation~\ref{eq:model}. For each SN, the parameters are the
 normalization and coordinates along the principal components $(x_k)$,
 a color and re-calibration parameters for spectra if any.  The actual
 components $(\M_k)$ and the parameters of the color-law $CL(\lambda)$
 also have to be estimated.  This procedure requires a first guess for
 the model components $(\M_k)$, for a first estimate of normalization,
 spectra re-calibration and color.  We used the SALT model SED
 sequence for a SN with stretch $=1$ for $\M_0$ and the difference of
 SED sequence of a SN with stretch $=1.1$ and the previous one for
 $\M_1$ (i.e. a linearized version of SALT model). Additional
 components where initiated with the orthogonal part of the SALT model
 SED sequence with respect to all previous components.

We end up with more than 3000 parameters to fit, with obvious
non-linearities, so that we used the Gauss-Newton procedure, which
consists in:
\begin{enumerate}
 \item Approximating locally the $\chi^2$ by a quadratic function of
 the parameters.
 \item Solving a large linear system to get an increment of the
   parameters $(\delta P_i)$.
 \item Increment the parameters and iterate until the $\chi^2$
   decrement with respect to the previous iteration becomes
   negligible.
\end{enumerate}

First, the average model is estimated along with the color-law,
calibration coefficients for spectra, and parameters of the SNe
( ($x_i$), $c$ ). When the system has converged, we add another
component, and all the parameters are fitted again (components,
color-law, SN parameters). The convergence algorithm is insensitive to
the input set of components.

\subsection{Regularization}

There might be some degeneracy in part of the phase space for the
 given data set.  For instance, if a phase$\times$wavelength region is
 only covered by photometry and not spectroscopy, we do not have
 enough data to constrain the combinations of parameters that model
 spectral features, whereas we can still model a photometric
 measurement, since the signal is integrated on a large spectral band.
 Adding a regularization term in the $\chi^2$ solves this issue. If
 its contribution is low enough, it will not alter significantly the
 determination of parameters that are addressed by the data, while
 putting some limitation on the parameters that are not.  We have
 chosen to minimize second derivatives with respect to phase and
 wavelength (once again, effective only when there is not enough
 data). The regularization term is the following :
$$
\chi^2_{REGUL} = n \times \sum { \M_k^T \D^T \D \M_k } \nonumber
$$ 
where $\M_k$ is the vector describing component $k$, $\D$ is the
derivative matrix and $n$ a normalization that controls the weight of
this regularization with respect to data.  Since such a term
introduces a bias in the estimator (departure from the maximum
likelihood estimator), we have to quantify it in order to adjust the
normalization $n$.  For this purpose, we used a simulated
dataset. This simulation helps us to define the resolution of the
model.  Each SN of the training sample was adjusted using the SALT
model, then fake light-curves and spectra were computed by replacing
each true measurement of the SN by the best fit value of the model.
The training procedure applied to this data set gives a result that is
slightly biased due to the regularization term in the $\chi^2$ in the
UV wavelength region.  The weight of the regularization term
(normalization $n$) was chosen so that the bias in $K$-corrections is
smaller than 0.005 mag for all wavelength, which is significantly less
than the statistical uncertainties.

\subsection{Model resolution}
The choice of the model resolution is imposed by the data set we have.
We used $10 \times 120$ parameters for $\M_0$ (10 along the time axis
and 120 for wavelength), in a phase range of $[-20,+50]$ days and a
spectral range of $[2000,9200]~\AA$.  This gives a spectral resolution
of order of $60~\AA$ which is sufficient for the modeling of SNe with
broad lines due to the velocity of the ejecta.  For $\M_1$, we choose
to use a lower resolution ( $10 \times 60$ parameters). The time axis
is remapped so that the time resolution at maximum and $+20$ days
after maximum is a factor two better than at $-20$ and $+50$ days
(approximately $4.5$ and $9$ days respectively).  As in SALT, we used
only two free coefficients to model the color law $CL(\lambda)$ (third
order polynomial, with two coefficients fixed so that
$CL(\lambda_B)=0$ and $ CL(\lambda_V)= 0.4 \log(10)$, see
Eq.~\ref{eq:model}). Using this number of parameters, when the model
is trained with the simulated data set described above, we found that
the limited resolution introduces a scatter in colors of only 0.01 mag
standard deviation (it is a scatter rather than a systematic effect
because of the varying epochs of photometric observations), which has
a negligible on distances when compared to the intrinsic dispersion of
SNe~Ia luminosities.

\section{Result of the training}
\label{sec:results}
We decided to consider only two components for the current analysis
since additional components are poorly constrained in most of the
phase space and marginally significant in the region of good data
coverage.  As the data sets improve so will the power to extract
additional components.  As a consequence, for each SN, we ended up
with four parameters, a date of $B-$band maximum, a normalization, the
parameter \1 and a color. The average value of \1 and its scale are
arbitrary since we can modify the components in consequence.  We
adopted $<x_1>=0$ and $<x_1^2>=1$.

Figure~\ref{fig:lc_templates} shows the variation of the UBVRI
light-curves as a function of parameter \1. We find that most of the
variability can be described by a simple stretching of light-curves
despite the fact that we did not force such behavior in the model.
More quantitatively, the parameter \1 can be converted into a stretch
factor~\footnote{Of course, since the model is a linear combination of
two components, we do not retrieve exactly the stretch model.}, whose
actual value depends on the reference light-curve template used, here
the one of SALT and of \citet{Goldhaber01} ($B$-band light curve
template ``Parab -18'', G01); or into $\Delta
m_{15}$~\citep{Phillips93} using the following transformations:
% HOWTO : drawsupermodel2 -1 -2 2 -e -r -15 +45 -E -u
\begin{figure}
\centering
\includegraphics[width=0.5\textwidth]{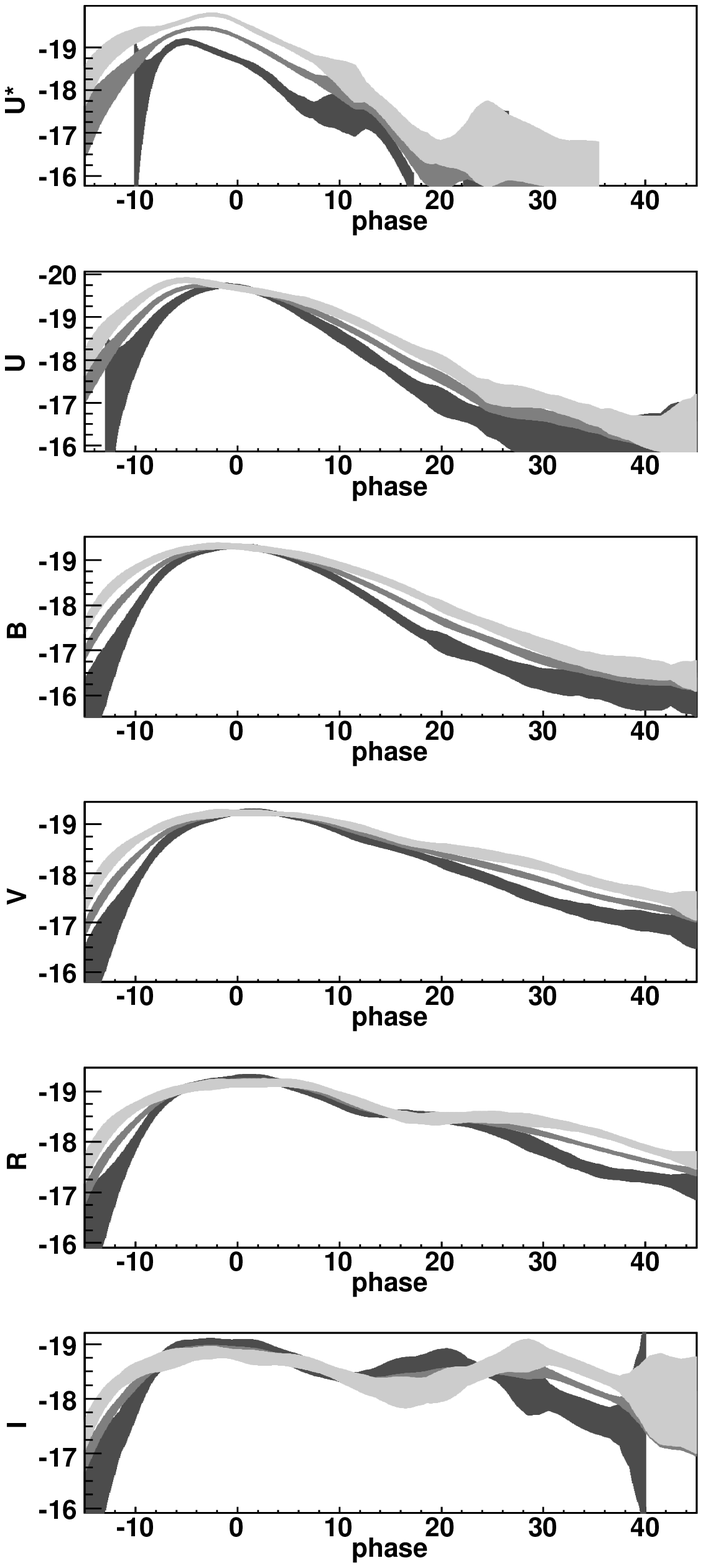}
\caption{The $U^*UBVRI$ template light curves obtained after the
training phase for values of \1 of -2, 0, 2 (corresponding to stretches of 0.8, 1.0, and 1.2; dark to light curves) and
null $B_V$ color excess. $U^*$ is a synthetic top hat filter in the
range 2500--3500~$\AA$.  The shaded areas correspond to the one
standard deviation estimate as described in section~\ref{sec:errors}.
\label{fig:lc_templates}} 
\end{figure}

% HOWTO : deltam15 ; l2root dm15.list ; root ~/salt2/work/macros/dm15.C
\begin{eqnarray}
& s\;(SALT) &= 0.98 + 0.091 x_{1} + 0.003 x_{1}^2 - 0.00075 x_{1}^3  \nonumber \\
& s\;(G01)  &= 1.07 + 0.069 x_{1} - 0.015 x_{1}^2 + 0.00067 x_{1}^3 \nonumber \\
& \Delta m_{15} &= 1.09 - 0.161 x_{1} + 0.013 x_{1}^2 - 0.00130 x_{1}^3 \nonumber
\end{eqnarray}

 Since there is not a perfect match of the non-linear stretch and
$\Delta m_{15}$ models with this one, those transformations (obtained
with simulations) vary with the weight attributed to each phase (the
scatter for stretch is about 0.02).

\begin{figure}
\centering
\includegraphics[width=0.5\textwidth]{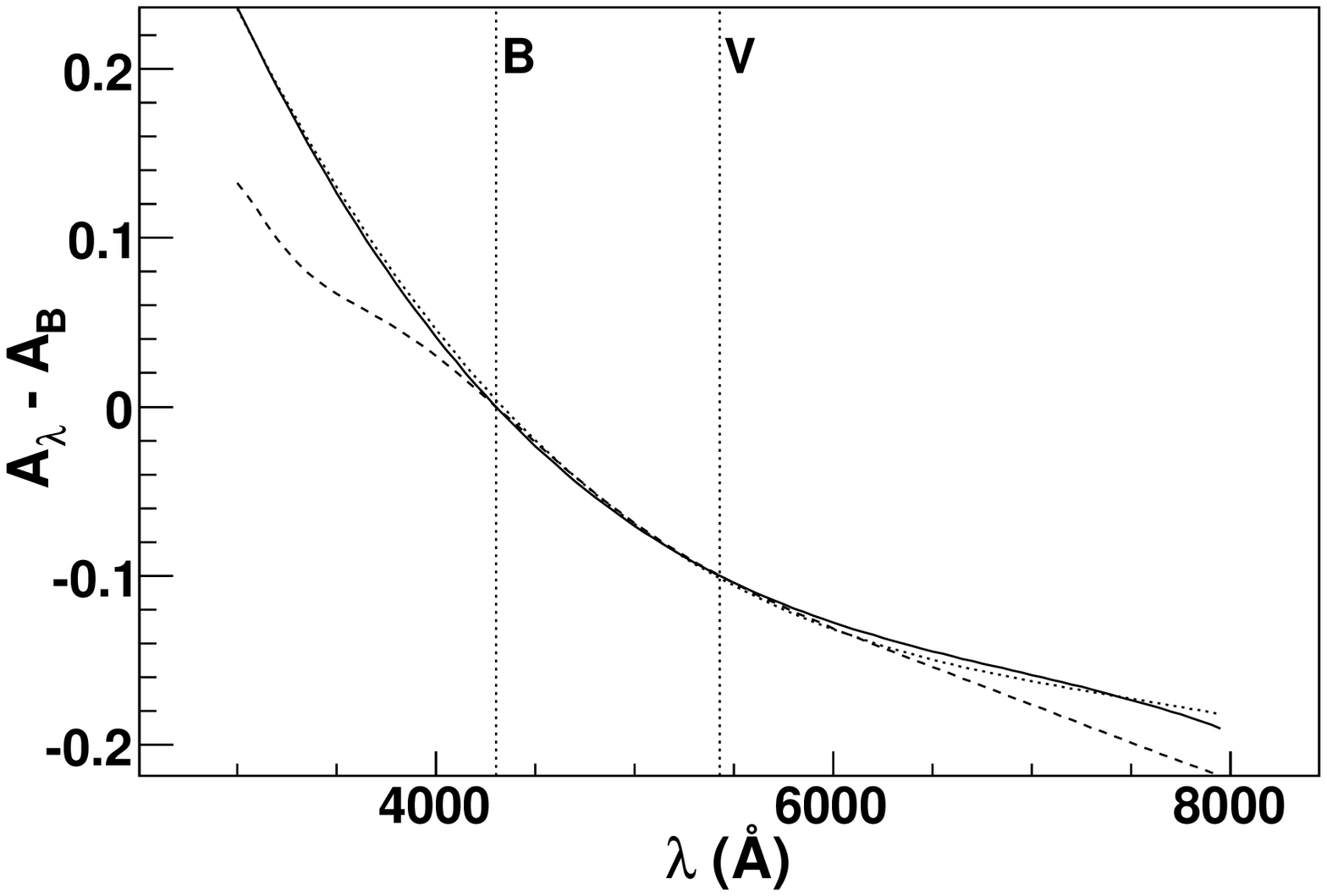}
\caption{The color law $c \times CL(\lambda)$ as a function of
wavelength for a value of $c$ of 0.1 (solid line). The dashed curve
represents the extinction with respect to $B$ band,
$(A_{\lambda}-A_B)$, from~\citet{Cardelli89} with $R_V=3.1$ and
$E(B-V)=0.1$, and the dotted line is the color law obtained with SALT
(very close to the result obtained here).
\label{fig:colorlaw}} 
\end{figure}

% HOWTO : ub ; l2root ub.list ; root ub.C
We also notice the $U-B$ color variation as a function of $B$-band
light-curve broadening.  The value of $U-B$ for phase=0 does not vary
with \1 (see Fig.~\ref{fig:lc_templates}), but when the flux is
integrated in the phase range -10, +10 days, we find that $U-B \propto
-0.2 \times s(SALT)$.  This is about half the value obtained with
SALT.  However, we find a color law very close to the one obtained
with SALT (see Fig.~\ref{fig:colorlaw}), despite the fact that the
supernova model is significantly different and the training set much
larger.

% HOWTO : drawextinction2 0.1

\begin{figure}
\centering
\includegraphics[width=0.5\textwidth]{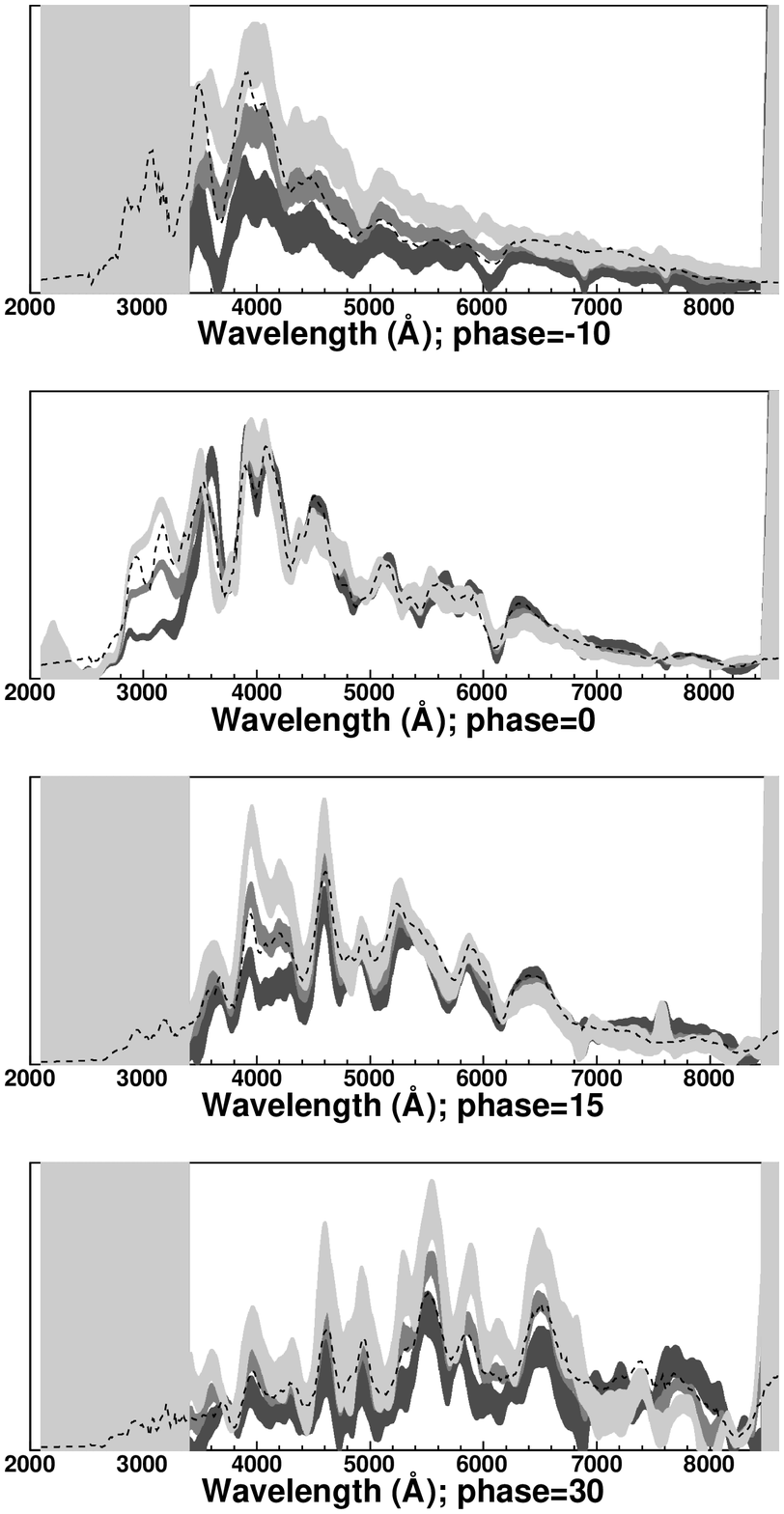}
\caption{Spectra at -10, 0, +15 and +35 days about B-band maximum for
values of \1 of -2, 0, 2 (corresponding to stretches of 0.8, 1.0, 1.2; light gray, black curve, dark gray) and null
$B_V$ color excess.  The shaded areas correspond to the one standard
deviation estimate as described in section~\ref{sec:errors}.  The
dashed curves represents the spectra of N02 (version 1.2).  All
spectra are arbitrarily normalized.
\label{fig:spectral_templates}} 
\end{figure}

\begin{figure}
\begin{center}
\includegraphics[width=\linewidth]{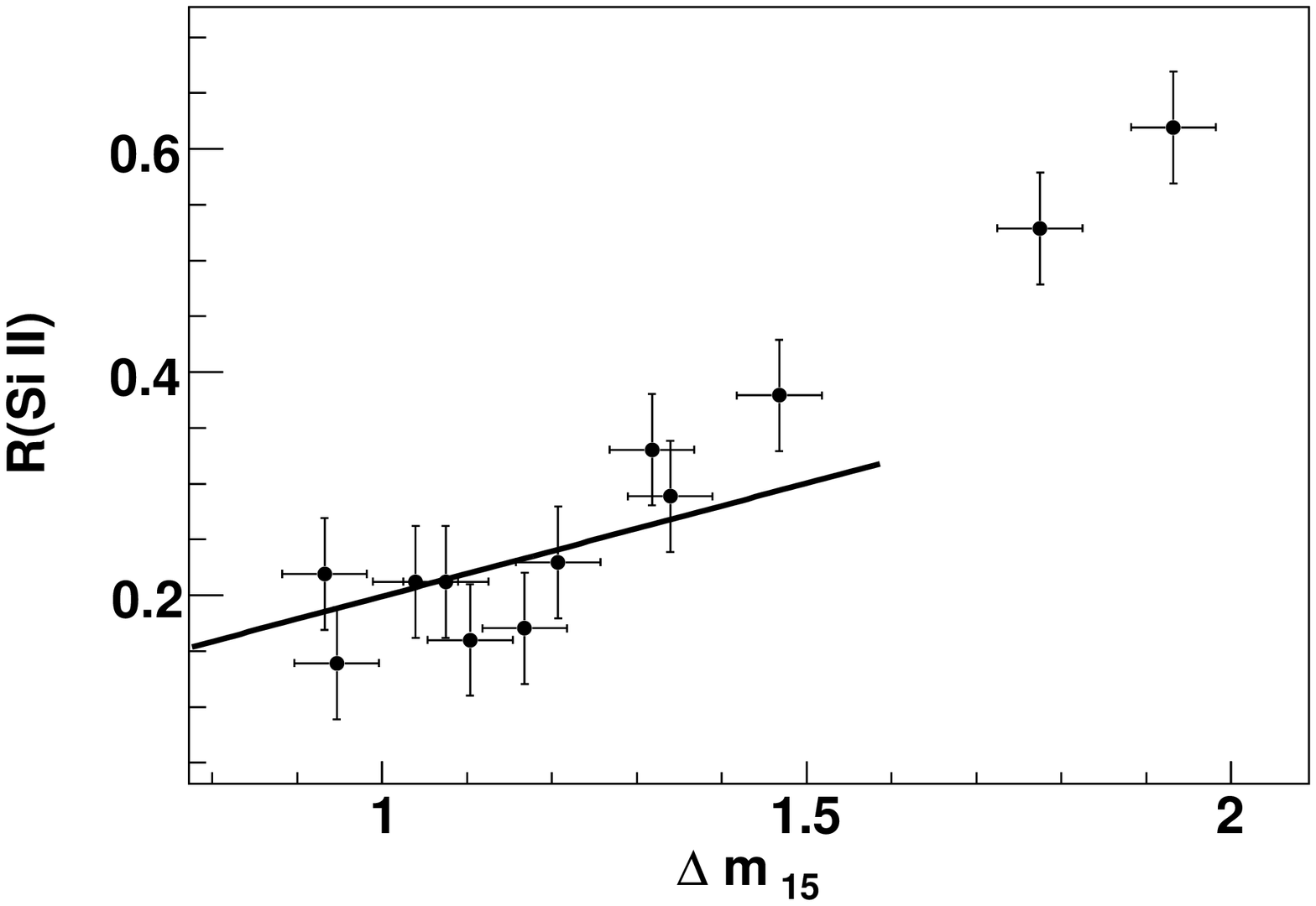}
\caption{The solid curve presents the relation of
R(Si~II)~\citep{Nugent95} as a function of $\Delta m _{15}$ along with
measurements from \citet{Benetti04}.
\label{fig:rs2}}
\end{center}
\end{figure}

These results confirm the main findings of SALT.  More interesting is
the variability of spectra with the first component as displayed in
Figure~\ref{fig:spectral_templates} for three phases about maximum.
It appears that the variability in U-band at maximum identified with
light-curves can be attributed to a sharp variation of the spectrum
for wavelengths lower than 3400~$\AA$.  It is possible to identify
such a feature thanks to the high redshift SNLS spectra.  Indeed, the
calibration of ground-based spectroscopic observations in the near
ultraviolet may not be reliable, whereas UV spectra obtained by the
International Ultraviolet Explorer have a very low signal to noise
ratio for wavelengths larger than 3000~$\AA$.  This feature can hardly
be approximated by a broad-band color evolution.  Hence we expect a
net improvement of the accuracy of distance estimates in the UV range
with respect to SALT or other equivalent methods which rely on a
single spectral sequence.  As an example of the modeling of the
variability of spectral features, figure~\ref{fig:rs2} presents the
variation of the R(Si II), as defined in \citet{Nugent95}, as a
function of $\Delta m_{15}$ retrieved from the model.  It is compared
to a compilation of observations by \citet{Benetti04}. There is a good
match in the $\Delta m_{15}$ range of the training sample ($\Delta m_{15}<1.6$).

One must however evaluate carefully the statistical significance of
the trends in the model.  Statistical errors of this model rely on the
weighting applied to spectra, and are sensitive to the errors assumed
for photometric measurements which are not very secure for most nearby
supernovae. Hence this accuracy of the modeling must be evaluated with
distributions of residuals as described in the next paragraph.

% HOWTO : drawmodelspectrum

\section{The remaining variability}
\label{sec:residuals}
%\subsection{Photometric residuals}

In order to assert the predictability of the model we need an
independent data set that is not used in the training
procedure. However, since we do not have a large number of
measurements available, we resorted to a jackknife procedure: for each
SN, we trained the model using all SNe from
Table~\ref{tab:training_sample} but this one, and looked at residuals
of the SN measurements to the retrieved model, fitting only the
parameters concerning this particular SN, i.e. the date of maximum,
($x_{0,1}$), and color.

The residuals obtained by this method are a priori highly correlated,
and this correlation is difficult to estimate from first principles.
However we would like to use this information to extract some
intrinsic variability of the SNe, beyond the principal components of
the model, in order to weight data according to this variability in
addition to measurement errors.  We resorted to the following
simplification, using two kinds of residuals (or model errors) : i)
Diagonal errors of the model are estimated from fits of light-curves
with an independent normalization for each of them. The residuals
obtained are of course still correlated, but we allow ourselves to
treat them as independent, assuming that the correlation length (along
time axis) is smaller than the data sampling for most
light-curves~\footnote{It is actually the purpose of the principal
component analysis to extract all the correlations between observables
for a given SN}.  ii) $K$-correction uncertainties, which can be
estimated using the difference between the peak magnitude obtained
from a single light-curve fit, and the one predicted by the model in
the same filter, fitting all light-curves.

\subsection{Diagonal uncertainties}
\label{sec:errors}
We use the correlations between the estimates of the components given
 by the correlation matrix retrieved at the end of the training
 procedure.  However, we allow ourself to scale these statistical
 uncertainties on the model in $p,\lambda$ bins to account for the
 remaining variability.

The model variance can then be defined as : 
\begin{eqnarray}
V_{MODEL}(x_1,p,\lambda) &=& S(p,\lambda) \times V_{MEAN}(x_1,p,\lambda) \nonumber \\ 
V_{MEAN}(x_1,p,\lambda) &=& \Hv_0^T \V_0 \Hv_0 + x_1^2  \Hv_1^T \V_1 \Hv_1 + 2 x_1  \Hv_0^T \C_{0,1} \Hv_1 \nonumber
\end{eqnarray}
where $\Hv_0(p,\lambda)$ and $\Hv_1(p,\lambda)$ are the vectors
defined in Sec.~\ref{sec:implementation} for components 0 and 1,
$\V_0$, $\V_1$ and $\C_{0,1}$ are the full variance and covariance
matrices of the components, and $S(p,\lambda)$ the scaling function.

In each $p,\lambda$ bin, $S(p,\lambda)$ is evaluated so that 
\begin{equation}
 \frac{1}{n} \sum \frac { \left[ f_i - x_0 \left(\Hv_{0,i}^T \M_{0} + x_1 \Hv_{0,i}^T \M_{1}\right) \right]^2}{ \sigma_i^2 + x_0^2 S(p,\lambda) V_{MEAN}(x_1,p,\lambda) } = 1 \nonumber
\end{equation}
where $(f_i)$ and $(\sigma_i)$ are the measurements and their
associated uncertainties.  We evaluated separately these errors for
light-curves and spectra, in order to take into account the correlated
errors along the wavelength axis when dealing with photometric data.

\begin{figure}
\centering
\includegraphics[width=0.5\textwidth]{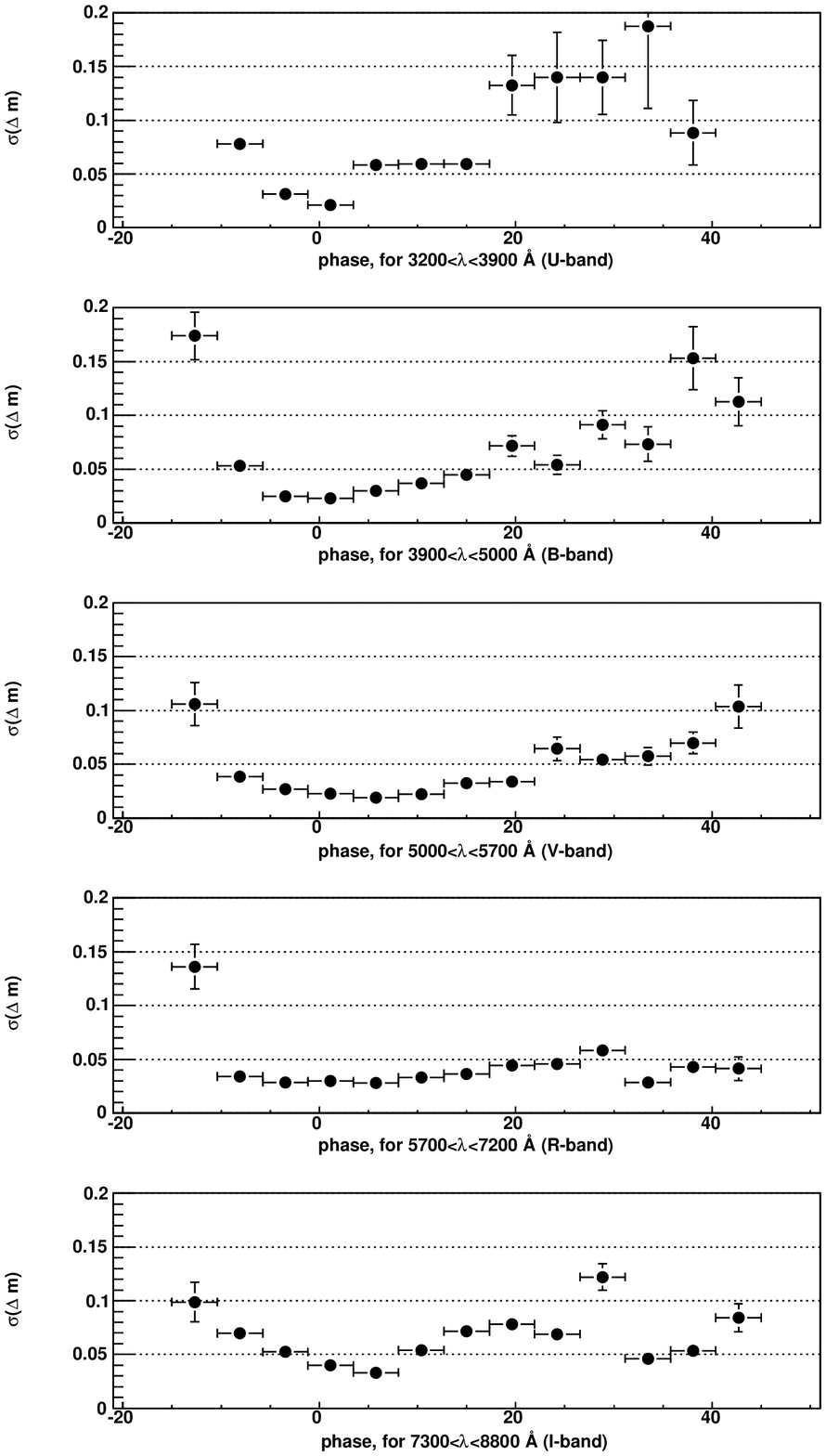}
\caption{Estimated standard deviation of model photometric errors as a
function of phase, for several rest-frame wavelength ranges roughly
corresponding from top to bottom to $U$, $B$, $V$, $R$ and
$I-$bands. Those model errors were evaluated from the scatter of
residuals to the single light-curve fit.
\label{fig:lc_residuals}} 
\end{figure}

Photometric residuals of the jackknife-like procedure are shown
figure~\ref{fig:lc_residuals}. The data set are split in four
rest-frame wavelength ranges, $[3200,3900]$, $[3900,5000]$,
$[5000,5700]$ and $[5700,7300]$ $\AA$, that roughly correspond to U,
B, V, R and I bands respectively. The model uncertainties basically
follow the statistical errors of data, with very good accuracy at peak
brightness and a poor quality at early and late phases, especially in
the $U-$band. In the rise time region, the large errors are partly due
to the limiting resolution of the model. Those errors are also
displayed in figure~\ref{fig:lc_templates} for $|x_1|$=2.

When fitting spectra, we imposed the values of the date of maximum,
$x_1$ and color obtained with the light-curves fit.  The only
remaining free parameters were those used to photometrically
"re-calibrate" spectra.  We found that model accuracy is poor at early
phases and in the UV region (Fig.~\ref{fig:spectral_templates}).

These estimates of the model errors can be accounted for when fitting
the light-curves or spectra. It gives more reliable statistical errors
for parameters (peak brightness, color, \1 and date of maximum) than
when only statistical errors of measurements are considered.

% HOWTO : 

\subsection{$K$-correction uncertainties}
\begin{figure}
\centering
\includegraphics[width=0.5\textwidth]{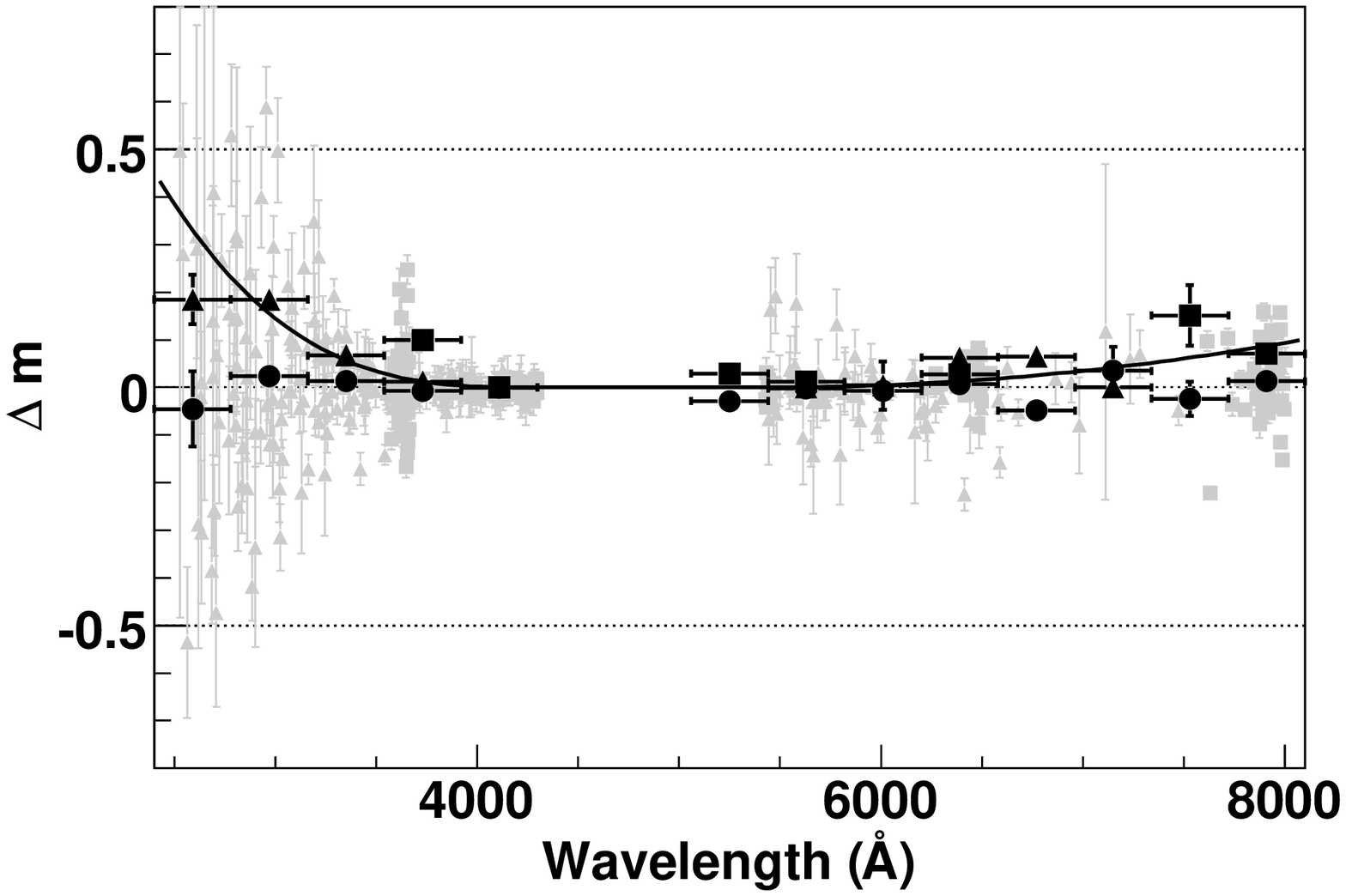}
\caption{Difference between observed peak magnitude in each band of
each SN from table~\ref{tab:training_sample} and the model prediction
as a function of the rest-frame effective wavelength of the filter
used (gray triangles : SNLS SNe, gray squares : nearby SNe). The large
black symbols represent the estimated dispersion in each wavelength
bin (triangles for SNLS, and squares for nearby SNe). The large
circles show the average difference in each wavelength bin for all SNe
and the solid curve is a polynomial fit to the dispersion used as an
estimate of the $K$-correction scatter. Since uncertainties on $B$ and
$V$ magnitudes at maximum enter in the normalization and color
evaluation of the model, $K$-correction uncertainties are set to zero
for $B$ and $V-$ band wavelengths.
\label{fig:kcorr}} 
\end{figure}

A direct approach to access the quality of $K$-corrections is to
compare the observed peak magnitude of a light-curve in a given filter
with the one predicted by the model using a fit of the other
light-curves.  Figure~\ref{fig:kcorr} presents the differences between
the observed and predicted magnitudes as a function of the effective
rest-frame wavelength of the instrument response used.

A more elaborated approach consists in modeling $K$-correction errors
with a parametric function of wavelength which value vanishes for
wavelength corresponding to the rest-frame $B$ and $V-$bands (errors
on $B$ and $V$ magnitudes at maximum enter in the normalization and
color evaluation). For each SN with enough light-curves, those
$K$-correction additional parameters can be estimated and their
standard deviation used to derive a model of $K$-correction
errors. Such a model is represented by the solid line
figure~\ref{fig:kcorr} and is given by the following formula:
\begin{eqnarray}
\sigma_{K} ( \lambda ) =   & 0.022 \left( \frac{\lambda - \lambda_B}{\lambda_U - \lambda_B} \right)^3  & \mathrm{ for } \, \lambda<\lambda_B \nonumber \\
 = & 0.018 \left( \frac{\lambda - \lambda_V}{\lambda_R - \lambda_V} \right)^2 & \mathrm{ for } \, \lambda>\lambda_V \nonumber
\end{eqnarray}

Since the estimate of $\sigma_{K}$ is based on the fit of the
normalization of light-curves, it measures a dispersion of colors
averaged over the phase range defined by the data set. Clearly, the
$K$-correction errors are large in the UV range. Also, those errors
must be added to the statistical errors on normalizations, but they do
not account for the whole observed scatter. For instance, for
$\lambda_U \simeq 3600 \AA$, we find a dispersion of 0.04 magnitude
(consistent with the one obtained in A06, Fig. 11.), but an
uncertainty of only 0.022 magnitude has to be added to the statistical
errors to match the observed dispersion. Diagonal uncertainties and
$K$-correction uncertainties are taken into account in the fits
performed in the following section.

% HOWTO : 
% superesiduals.csh 
% cd work/list ; kcorrection_error_fit kcorrections.list
% l2root kcorr.list ; l2root kcorr2.list ; root ../macros/kcorr.C 

\section{Improving the distance estimates of distant SNe}
\label{sec:improving}

Despite the fact that our modeling is less accurate in the UV range,
it is still very useful for distance estimates of high-redshift SNe
($z>0.8$) for which, as in the case of SNLS, the rest-frame $B$ and
$V$-band observations often have a very poor signal to noise ratio.

\subsection{Light-curve fit of distant supernovae}

% HOWTO :
% snfit lc2fit_{g,r,i,z}.dat -w 2000 9000 -e -k -m salt2 |& tee snfit_salt2.log
% TOT chi2/dof 0.90387141062 (NDF:53)
% drawlc lc2fit_{g,r,i,z}.dat -w 2000 9000 -p result_salt2_griz.dat -e -L SNLS-04D3gx -k

\begin{figure}
\centering
\includegraphics[width=\linewidth]{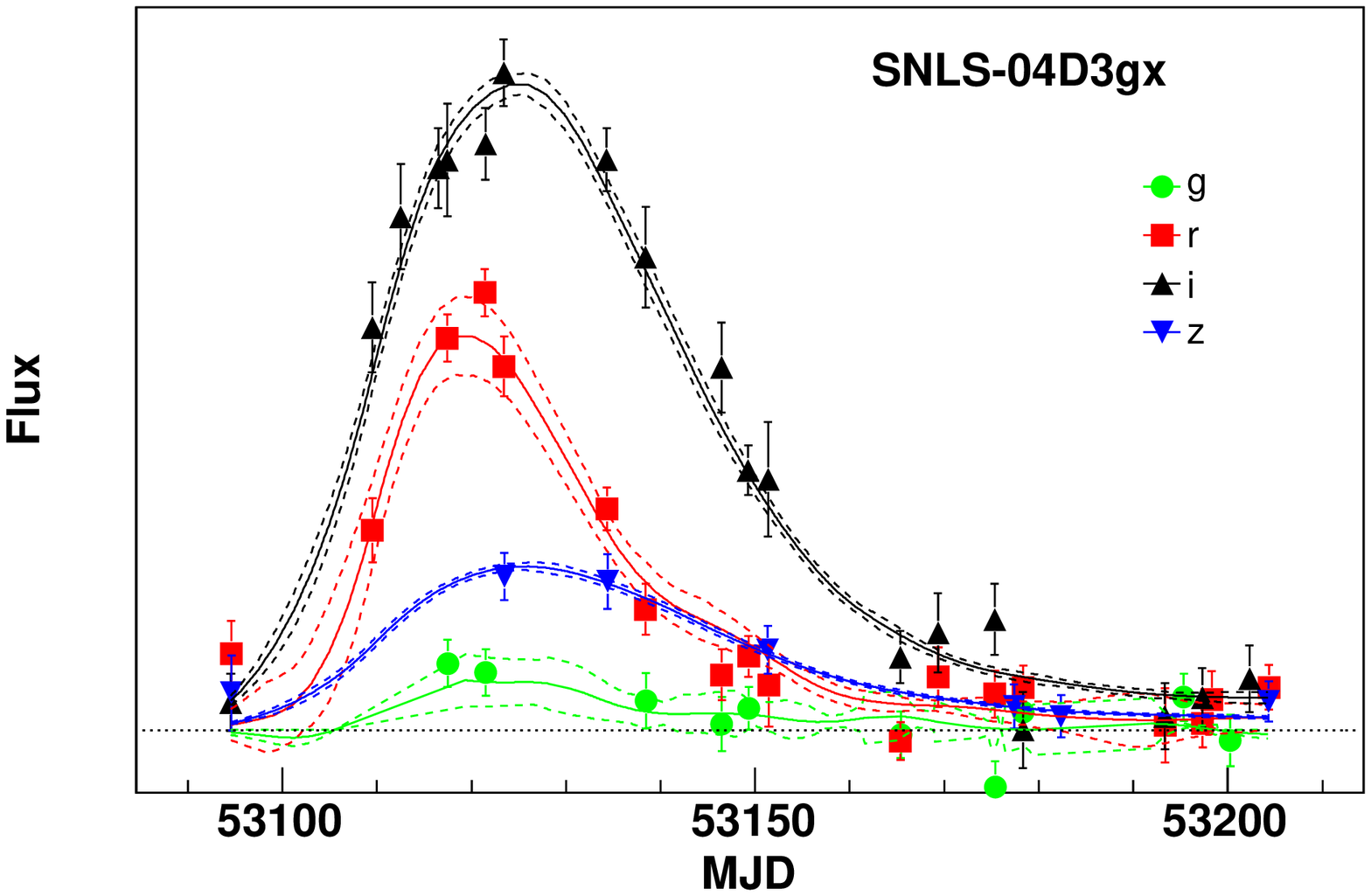}
\caption{Observed light-curves points of the SN~Ia SNLS-04D3gx at
z=0.91 along with the light-curves derived from the model (solid line,
trained without this SN).  The dashed lines represent the 1 $\sigma$
uncertainties of the model (both uncorrelated and $K$-correction
errors).
\label{fig:04D3gx}}
\end{figure}

Figure~\ref{fig:04D3gx} shows the SN~Ia SNLS-04D3gx at $z=0.91$ fitted
by the model.  All four light-curves (g,r,i,z) are well described by
the best fit model for which only four parameters were adjusted (date
of maximum, normalization, color and \1). The $\chi^2$ per degree of
freedom (d.o.f.) of the fit is 0.76 (for 50 d.o.f.) when diagonal and
$K$-correction uncertainties of the model are considered.
Table~\ref{tab:color_errors} illustrates the gain in the accuracy of
the color estimate for SNLS-04D3gx.

\begin{table}[h]
%iz -k -e   Color -0.229165857057 0.180059633719
%iz -e      Color -0.229165886114 0.1800
%iz         Color -0.224868888384 0.176878
%riz -k -e  Color -0.14746866651 0.0756723863958
%riz -e     Color -0.13818529311 0.0576268980028
%riz        Color -0.15022128461 0.0514824020371
%griz -k -e Color -0.172359386007 0.0703003541835
%griz -e    Color -0.168871187855 0.0519597476557
%griz       Color -0.174986584946 0.0468022127783
\begin{tabular}{lll}
bands & $\lambda_{\mathrm{min}}$ & color\\ 
\hline
i,z     & 3980 & $-0.220 \pm 0.180\mathrm{(s)} \pm 0.033\mathrm{(d)}\pm 0.005\mathrm{(k)}$ \\
r,i,z   & 3250 & $-0.147 \pm 0.051\mathrm{(s)} \pm 0.026\mathrm{(d)}\pm 0.049\mathrm{(k)}$ \\
g,r,i,z & 2520 & $-0.172 \pm 0.047\mathrm{(s)} \pm 0.023\mathrm{(d)}\pm 0.047\mathrm{(k)}$
\end{tabular}
\caption{Error on the color estimate of SNLS-04D3gx as a function of
the number of light-curves included in the fit. The contributions to
the error are measurement statistical errors (s), diagonal model
errors (d) and $K$-correction errors (k).
\label{tab:color_errors}}
\end{table}

The total uncertainty on the color parameter is reduced by a factor
2.5 when g and r-band (rest-frame UV) light-curves are included in the
fit. We see that the model uncertainties are large in this wavelength
range and therefore must be propagated.

\subsection{Improving cosmological results}
\label{sec:cosmology}

The parameters retrieved from the light-curve fit can be used to
estimate distances using the same procedure as described in A06. The
distance estimator is a linear combination of $m^{*}_B, x_1$ and $c$ :
$$
\mu_B = m^{*}_B - M +\alpha_x \times x_1 - \beta \times c
$$ 
with $m^{*}_B$, $x_1$ and $c$ derived from the fit to the light
curves, and $\alpha_x$~\footnote{Note that the definition of
$\alpha_x$ differs from that of $\alpha$ in SALT}, $\beta$ and the
absolute magnitude $M$ are parameters which are fitted by minimizing
the residuals in the Hubble diagram.  As in A06, we introduce an
additional "intrinsic" dispersion ($\sigma_{int}$) of SN absolute
magnitudes to obtain a reduced $\chi^2$ of unity for the best fit set
of parameters.

We minimize the following functional form, which gives negligible biases to the 
estimates of $\alpha_x$ and $\beta$: 
$$
\chi^2 = \sum_s \frac{ {\bf V}^T {\bf X}_s - M - 5 \log_{10} \left( d_L({\bf \theta},z)/ 10 \, \mathrm{pc} \right)} {\bf{V}^T \; {\bf C}({\bf X}_s) \; {\bf V}}
$$
with
$$
{\bf X}_s = 
\begin{pmatrix}
m^{*}_B \\
x_1 \\
c \\
\end{pmatrix}
, \,\, 
{\bf V} = 
\begin{pmatrix}
1 \\
\alpha_x \\
-\beta \\
\end{pmatrix} 
$$ ${\bf \theta}$ stands for the cosmological parameters that define
the fitted model and $d_L$ is the luminosity distance.  ${\bf C}({\bf
X}_s)$ is the covariance matrix of the parameters ${\bf X}_s$ for
which we have included in the variance of $m^{*}_B$ the intrinsic
dispersion and an error of the distance modulus due to peculiar
velocities, which we take to be 300 km.s$^{-1}$.

To estimate the systematic effect due to modeling of the SN~Ia SED sequence, 
we fit the data set of A06, which consists in 44 nearby SNe~Ia and  71 SNLS SNe.
Whereas we used improved photometry for the training of the model, we consider here
exactly the same data set as in A06, in order to ease the comparison with the results obtained
with SALT.

\begin{figure}
\centering
\includegraphics[width=0.5\textwidth]{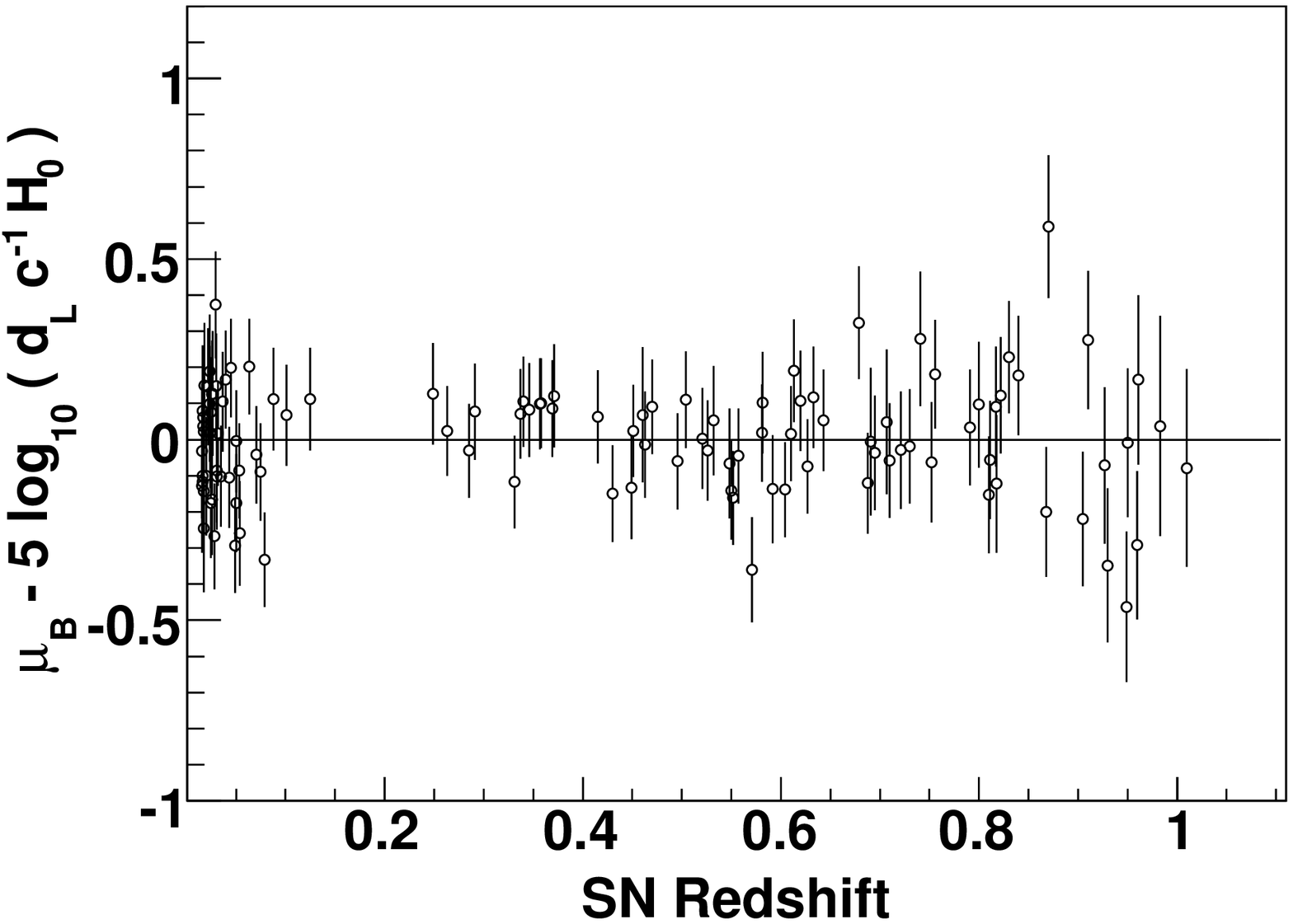}
\caption{Residuals to the distance redshift relation for the best fit $\Lambda$CDM cosmology for SNLS and nearby SNe~Ia.
\label{fig:hubble}} 
\end{figure}

Thanks to the evaluation of the model errors, we can safely use all
available light-curves in the fit. Especially the $r-$band data at
redshifts greater than 0.8 (effective rest-frame wavelength lower than
3440~$\AA$) are very useful to constrain the color of the supernova.
With this additional information, the uncertainty in distance moduli
is significantly reduced at high redshift, yielding a better
resolution on cosmological parameters.  For flat $\Lambda$CDM
cosmology, we obtain :
\begin{eqnarray}
\om & = & 0.240 \pm 0.033 \nonumber \\
\alpha_x & = &  0.13 \pm 0.013 \nonumber \\
\beta & = & 1.77 \pm 0.16  \nonumber
\end{eqnarray}
with $\sigma_{int}=0.12$ (which is smaller than the value of 0.13 in
A06 because we have considered uncertainties in the model).  The RMS
of the residuals around the best fit Hubble relation is 0.161 mag
(compared to 0.20 in A06, see Fig.~\ref{fig:hubble}).

The uncertainty on $\om$ is improved by 10\% with respect to the A06
($\om=0.263 \pm 0.037$, table 3 of~\citealt{Astier06}) analysis.  We
find a difference of 0.023 on $\om$ which is consistent with the
assumed systematic error due to modeling of 0.02 in that
paper. However the supernova SNLS-03D1cm at a redshift of 0.87 now
appears as a significant outlier as the distance modulus resolution
improved~\footnote{SNLS-03D1cm is $0.6 \pm 0.2 $ mag dimmer than
expected for the best fit cosmology, which corresponds to a 3 $\sigma$
deviation when including the intrinsic dispersion. This has a 27\%
probability to occur at least once by chance for our sample of 115
SNe, if SNe distances are scattered about the Hubble law following a
Gaussian distribution.}.  This object has a spectrum with low signal
to noise ratio and was classified as probable Ia.  Discarding this
object from the analysis gives $\om = 0.246 \pm 0.032$ and a standard
deviation of residuals of 0.154 magnitude.

Since the current model significantly improves distance estimates at
high redshifts, we obtain a greater improvement on the estimation of
the equation of state of dark energy. As an example, we may use the
figure of merit proposed by the Dark Energy Task Force~\citep{DETF06},
which is inversely proportional to the area of the 95\% confidence
level contour in the plane $(w_a,w_p)$, where the following
parametrization is considered for the equation of state of dark energy
:
$$
w = w_p + (a_p-a) w_a
$$ $a$ being the scale factor, and $a_p$ a reference scale factor
chosen so that the estimates of $w_p$ and $w_a$ are uncorrelated for a
given experiment. Using the baryon acoustic oscillations constraints
from \citet{Eisenstein05} (Eq. 4), $a_p = 0.851$, and we improve this
figure of merit by 35\% with respect to the analysis using SALT (for
which $a_p$ is slightly different).

Light-curves can be fitted with all the models from the jackknife
procedure, so that we can derive as many estimates of $\om$ as there
are SNe in the training sample. We found that the RMS of this
distribution is 0.003, so that we expect a deviation from a model
trained with an infinite number of SNe of the order of 0.0015 on $\om$
(for a flat $\Lambda$CDM cosmology), a value that is negligible
compared to the other sources of systematic errors in such an
analysis.

%HOWTO : drawcosmo -C 0.26 0.74 -2 -L work/lists/snfitresults_2005_2900_7000.list -a 0.14 -b 0 -g -1.7 -e 0.12 -M -19.3

\section{Other applications}
\label{sec:applications}

%%%%%
% A MODIFIER
%%%%%

The proposed model provides a tool for spectroscopic and photometric
identification, and photometric redshift determination.  A detailed
analysis of the purity and efficiency of a photometric identification
tool with respect to SNe~Ib, Ic and II is beyond the scope of this
paper.

\subsection{Spectroscopic identification}
\label{sec:spectroscopic_identification}

The model proposed allows simultaneous fits of light-curves and
spectra \New{(with additional galaxy templates to evaluate the host
contamination as mentioned in Sec.~\ref{sec:highz_sample})}.  This can
turn out to be very useful for identification of Type Ia supernovae.
Indeed, some Type Ic SNe can present light-curves and spectra that
look qualitatively like SNe~Ia, and in the most extreme cases,
photometric and spectroscopic identifications taken separately may
fail to tag this object correctly. A detailed analysis is beyond the
scope of this paper. We will show only two examples.

\begin{figure}
\begin{center}
\begin{minipage}[c]{\linewidth}
\includegraphics[width=\linewidth]{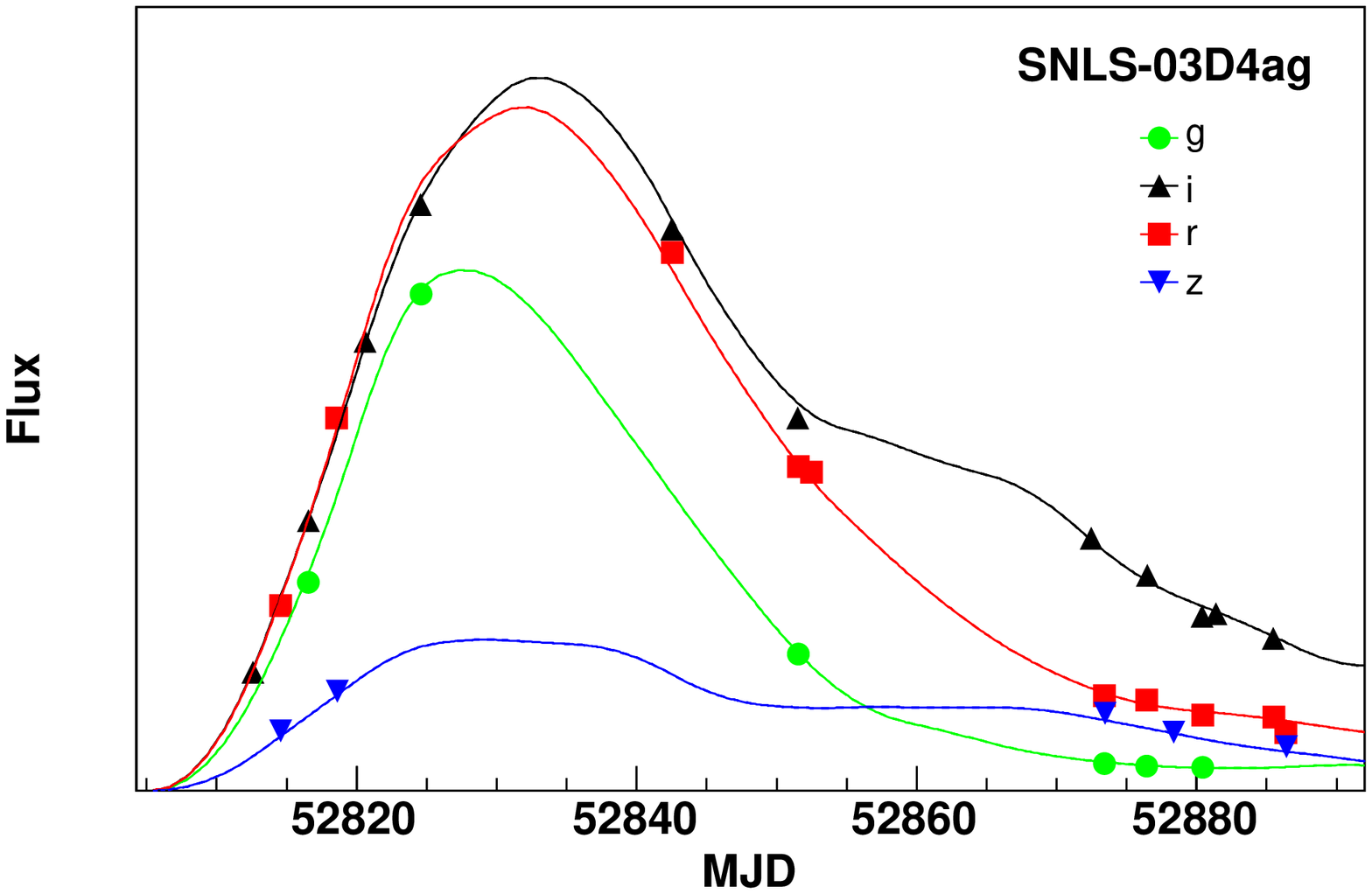}
\end{minipage}
\begin{minipage}[c]{\linewidth} 
\includegraphics[width=\linewidth]{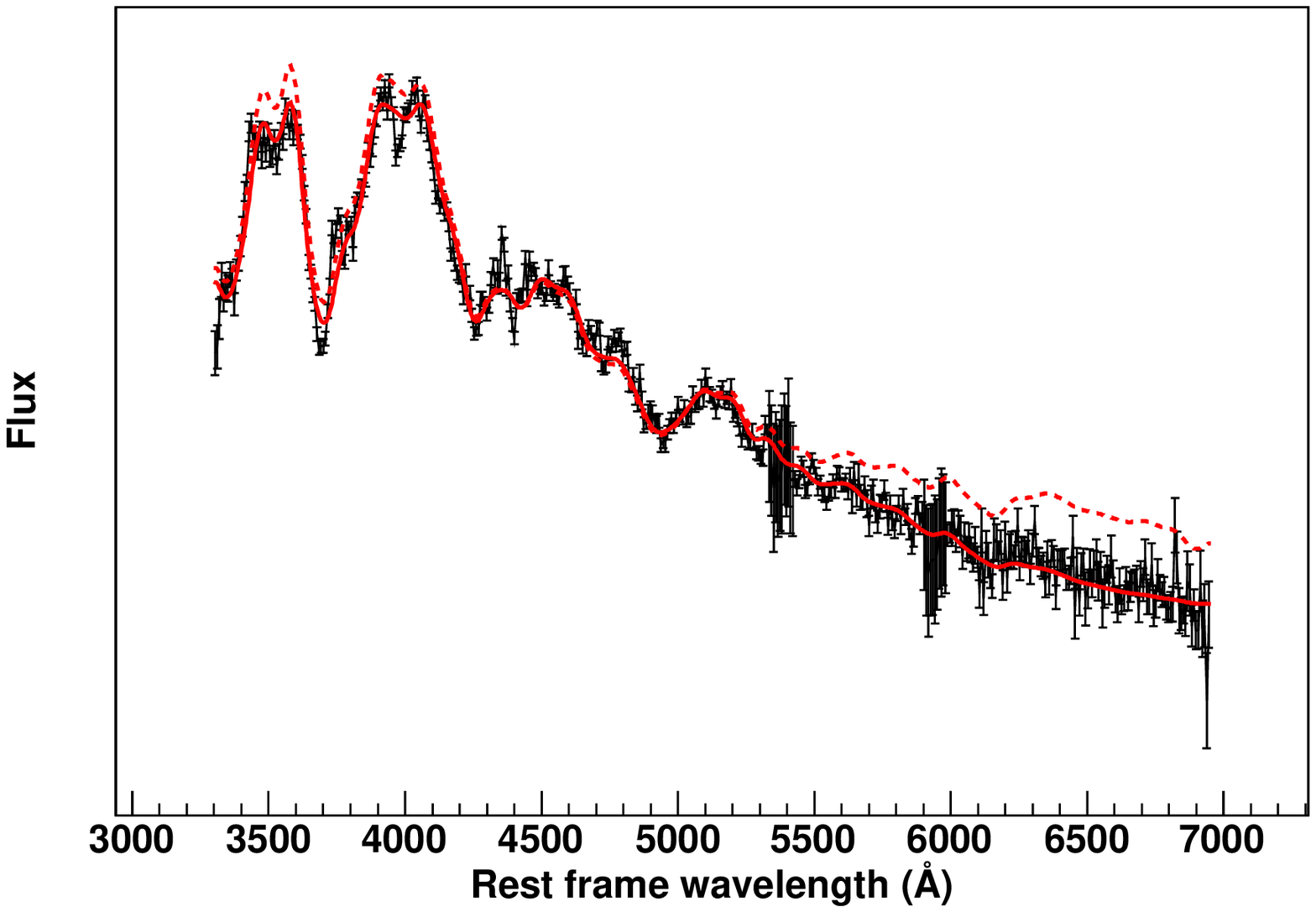}
\end{minipage}
\caption{Top: Observed light-curves points of the SN~Ia SNLS-03D4ag at
z=0.285 along with the light-curves derived from the model (solid
line). Bottom: Observed VLT spectrum at a phase of -8 days along with
the model spectrum. The dashed curve is the expected model from the
light-curve fit (up to a global normalization) and the solid curve the
one including four re-calibration parameters.
\label{fig:03D4ag}}
\end{center}
\end{figure}

First, figure~\ref{fig:03D4ag} shows the observed light-curves and
spectrum of SNLS-03D4ag at a redshift of 0.285 along with the result
of the simultaneous fit. Four ``re-calibration'' parameters for the
spectrum were considered (since there are four light-curves).  The
$\chi^2$ per degree of freedom for the light-curves and the spectrum
are respectively 0.53 (for 28 d.o.f.) and 0.63 (for 1770 d.o.f.),
taking into account the model errors~\footnote{\New{With the model
errors, the average $\chi^2$ per degree of freedom for all the SNe of
the training sample is one by definition}}), so that this SN can be
safely considered as a typical normal SN~Ia.

% HOWTO :
% snfit lc2fit_*.dat -m salt2 -k -e -w 2000 9000 -r -15 +45 | & tee snfit.log
% snfit -s spectrum_03D4ag_180_PSFspec_salt.dat -m salt2 -f z 0.285 -f daymax 52831.0448601 -f X1  0.72826110974  -f color  -0.0954865500 -c 4 -e | & tee snfit_spec.log
% All LC chi2/dof 
% All SPEC chi2/dof 
% drawlc lc2fit_?.dat -p result_salt2.dat -L SNLS-03D4ag
% drawspectrum spectrum_03D4ag_180_PSFspec_salt.dat -p result_salt2.dat -r -b 10
%

% HOWTO :
% snfit lc2fit_*.dat -m salt2 -k -e -w 2000 9000 -r -15 +45 | & tee snfit.log
% 
Now if we consider an SN~Ic, for instance SNLS-03D4aa at a redshift of
0.166, the model gives a very bad fit of the data (reduced $\chi^2$ of
4.6 (for 10 d.o.f.)  and 1.6 (for 1807 d.o.f) for light-curves and
spectra respectively), allowing for a clear rejection of this event.

%\begin{figure}
%\begin{center}
%\begin{minipage}[c]{\linewidth}
%\includegraphics[width=\linewidth]{03D4aa_lcs.eps}
%\end{minipage}
%\begin{minipage}[c]{\linewidth} 
%\includegraphics[width=\linewidth]{spectrum_03D4aa_180_PSFspec.dat.eps}
%\end{minipage}
%\caption{Top: Observed light-curves points of the SN~Ic SNLS-03D4aa at z=0.166 along with 
%the light-curves derived from the model (solid line). Bottom: Observed VLT spectrum at a phase of +2 days along with the model spectrum. The dashed curve is the expected model from the light-curve fit (up to a global normalization) and the solid curve the one including four re-calibration parameters. The model is obviously a very bad fit to the data. 
%\label{fig:03D4aa}}
%\end{center}
%\end{figure}

\subsection{Photometric redshifts}
\label{sec:photoz}

For photometric redshift determination, one can compare the redshift
estimate based on a simultaneous fit of all light-curves of a given
supernova and the much more precise spectroscopic redshift derived
from the spectroscopic observation of the host of the object. Using
all available light-curves (up to four for SNLS) allows us to relax
assumptions on the light curve shape parameter ($x_1$) and the color,
without any use of the absolute luminosity, so that we do not require
any prior on the cosmological parameters.

We have applied this method to all SNe~Ia from
Table~\ref{tab:training_sample} with at least 3 light-curves, using
for each SN the model obtained without this SN (always to avoid
over-training). A Gaussian prior for \1 was assumed based on the
statistics from the training procedure, $x_1 = 0 \pm 1$. No prior was
applied to color.

\begin{figure}
\centering
\includegraphics[width=0.5\textwidth]{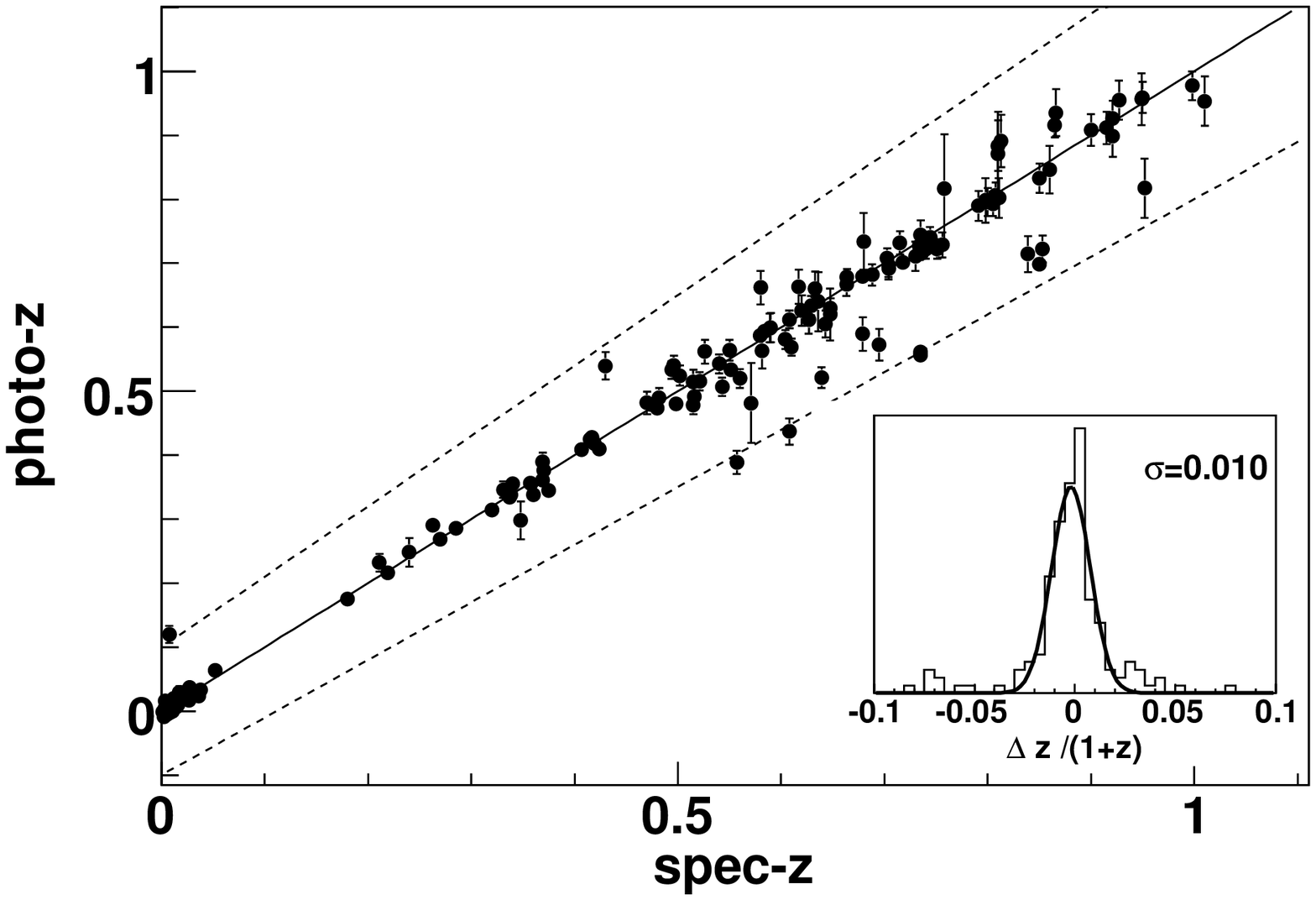}
\caption{Photometric vs spectroscopic redshifts for the training
 sample of SNe. The dashed lines are $z \pm 0.1 (1+z)$ functions.
\label{fig:photoz}} 
\end{figure}

The resulting photometric redshift is compared to the spectroscopic
one on Figure~\ref{fig:photoz}. The RMS of the distribution of
$\Delta z/(1+z)$ is 0.02 with no significant bias. A Gaussian fit of
the distribution of $\Delta z/(1+z)$ gives $\sigma=0.01$. This is an
accuracy of the same order of magnitude as the one that can be derived
from a fit of the SN spectrum alone.  The only low-redshift SN for
which the photometric redshift is off by more than 0.1 is SN1999cl
which is a highly extincted supernova.
%, so that the prior on color introduces a strong bias for a higher redshift.

% HOWTO : cd ~/pca ; source setup.csh ; photoz.csh ;  root macros/photoz.C

\section{Conclusion}
\label{sec:conclusion}

We have proposed a new empirical model of Type Ia supernovae
spectro-photometric evolution with time, based on observed nearby and
distant supernovae light-curves and spectra. The method uses
available spectra, regardless of their wavelength range or calibration
accuracy, since they are re-calibrated using the photometric
information in the training process.  This model provides an average
spectral sequence of Type Ia supernovae and their principal
variability components~\footnote{available together with the fitter source code at http://supernovae.in2p3.fr/$\sim$guy/salt}.

Thanks to an evaluation of the modeling errors (for photometric
 points, spectra, and broad-band colors), on can safely use most of
 the information on a given supernova for comparison with the model.
 This is very helpful for distance estimates, but also photometric
 redshift evaluation and SN identification.

Applied to the supernovae sample of A06, we improved the distance
estimates, especially at redshifts larger than 0.8, thanks to the
modeling of the ultraviolet emission and the propagation of model
errors.  When a flat $\Lambda$CDM cosmology is fitted to the Hubble
diagram, the gain in statistical resolution on $\om$ is of 10\%
compared to the previous analysis using SALT; it is 35\% better on the
two parameter constraint of the equation of state of dark energy.

The model accuracy is currently limited by the size and quality of the
 current data sample.  A larger sample would be needed to look for
 another variability component.

This empirical modeling technique is well suited for the analysis of
very large samples of supernovae photometric data such as the ones
expected from future projects like DUNE~\citep{Refregier06},
JDEM~\citep{AlderingSNAP} or LSST~\footnote{LSST:www.lsst.org}. 
Indeed, with large statistics of high precision 
photometric data spread on a large redshift range, we may expect to
deconvolve the SED sequence of the average supernova and its principal
variations. Such an analysis would permit one to look for variability
correlated with redshift, and hence provide a non-parametric approach
to test for supernova evolution.

\begin{acknowledgements}
We gratefully acknowledge the assistance of the CFHT Queued Service
Observing Team, led by P. Martin (CFHT). We heavily rely on the
dedication of the CFHT staff and particularly J.-C.~Cuillandre for
continuous improvement of the instrument performance. The real-time
pipelines for supernovae detection run on computers integrated in the
CFHT computing system, and are very efficiently installed, maintained
and monitored by K.~Withington (CFHT). We also heavily rely on the
real-time Elixir pipeline which is operated and monitored by
J-C.~Cuillandre, E. Magnier and K.~Withington. We are grateful to
L.~Simard (CADC) for setting up the image delivery system and his kind
and efficient responses to our suggestions for improvements. The
French collaboration members carry out the data reductions using the
CCIN2P3.  Canadian collaboration members acknowledge support from
NSERC and CIAR; French collaboration members from CNRS/IN2P3,
CNRS/INSU, PNC and CEA.
\end{acknowledgements}

%_____
\newpage
\bibliographystyle{aa}
\bibliography{bibi}

\clearpage
%% \pagebreak
\onecolumn
% this is the training sample
% generated by make_training_table.csh

\begin{deluxetable}{lcccccc}
\tablewidth{0pt}
\tablecaption{The training sample of Type Ia supernovae light curves and spectra}
\label{tab:training_sample}
\tablehead{
\colhead{Name} & 
\colhead{$z$ \tablenotemark{a}} & 
\colhead{Bands\tablenotemark{b}} &
\colhead{Number of spectra; phases\tablenotemark{c}}
}
\startdata
1981B & 0.006 & UBV~(B83) & 6 ; \, $ 0 \le p \le 35 $~(Br83)  \\
1986G & 0.002 & BV~(P87) & 6 ; \, $ -3 \le p \le 18 $~(AC)  \\
1989B & 0.002 & UBVR~(W94) & 18 ; \, $ -6 \le p \le 19 $~(AC)  \\
1990af & 0.051 & BV~(H96) & 0~  \\
1990N & 0.003 & UBVR~(L91,L98) & 15 ; \, $ -14 \le p \le 38 $~(AC,IUE,CFA)  \\
1991T & 0.006 & UBVR~(F92,L98,A04,K04b) & 36 ; \, $ -13 \le p \le 258 $~(AC,CFA)  \\
1992A & 0.006 & UBVR~(S92,A04) & 19 ; \, $ -5 \le p \le 27 $~(CFA,IUE)  \\
1992al & 0.015 & BVR~(H96) & 0~  \\
1992bc & 0.020 & BVR~(H96) & 0~  \\
1992bo & 0.019 & BVR~(H96) & 0~  \\
1993O & 0.052 & BV~(H96) & 0~  \\
1994ae & 0.004 & BVR~(R99,A04) & 0~  \\
1994D & 0.001 & UBVR~(R95,P96,M96,A04) & 21 ; \, $ -11 \le p \le 76 $~(AC)  \\
1994S & 0.015 & BVR~(R99) & 1 ; \, $ p = 22 $~(AC)  \\
1995ac & 0.050 & BVR~(R99) & 0~  \\
1995al & 0.005 & BVR~(R99) & 0~  \\
1995bd & 0.016 & BVR~(R99) & 0~  \\
1995D & 0.007 & BVR~(R99,A04) & 0~  \\
1995E & 0.012 & BVR~(R99) & 0~  \\
1996bl & 0.036 & BVR~(R99) & 0~  \\
1996bo & 0.017 & UBVR~(R99,A04) & 0~  \\
1996X & 0.007 & BVR~(R99) & 12 ; \, $ -1 \le p \le 57 $~(AC)  \\
1997bp & 0.008 & UBVR~(J05) & 0~  \\
1997bq & 0.009 & UBVR~(J05) & 0~  \\
1997do & 0.010 & UBVR~(J05) & 0~  \\
1997E & 0.013 & UBVR~(J05) & 0~  \\
1998ab & 0.027 & UBVR~(J05) & 0~  \\
1998aq & 0.004 & UBVR~(R05) & 26 ; \, $ -9 \le p \le 90 $~(CFA)  \\
1998bu & 0.003 & UBVR~(S99) & 66 ; \, $ -3 \le p \le 56 $~(CFA)  \\
1998dh & 0.009 & UBVR~(J05) & 0~  \\
1998ef & 0.018 & UBVR~(J05) & 0~  \\
1998es & 0.011 & UBVR~(J05) & 0~  \\
1999aa & 0.014 & UBVR~(J05) & 9 ; \, $ -11 \le p \le 40 $~(G04)  \\
1999ac & 0.009 & UBVR~(J05) & 0~  \\
1999aw & 0.038 & BVR~(St02) & 0~  \\
1999cl & 0.008 & UBVR~(J05,K00) & 0~  \\
1999dk & 0.015 & UBVR~(K01,A04) & 0~  \\
1999dq & 0.014 & UBVR~(J05) & 0~  \\
1999ee & 0.011 & UBVR~(S02) & 11 ; \, $ 0 \le p \le 43 $~(H02)  \\
1999ek & 0.018 & UBVR~(J05,K04b) & 0~  \\
1999gp & 0.027 & UBVR~(J05,K01) & 0~  \\
2000E & 0.005 & UBVR~(V03) & 0~  \\
2000fa & 0.021 & UBVR~(J05) & 0~  \\
2001ba & 0.029 & BV~(K04a) & 0~  \\
2001bt & 0.014 & BVR~(K04b) & 0~  \\
2001cz & 0.016 & UBVR~(K04b) & 0~  \\
2001el & 0.004 & UBVR~(K03) & 2 ; \, $ p = 24,33 $~(W03)  \\
2001V & 0.015 & BVR~(Vi03) & 0~  \\
2002bo & 0.004 & UBVR~(Z03,B04,K04b) & 12 ; \, $ -14 \le p \le 38 $~(B04)  \\
2002er & 0.009 & UBVR~(P04) & 0~  \\
2003du & 0.006 & UBVR~(A05) & 4 ; \, $ -11 \le p \le 37 $~(A05)  \\
2004fu & 0.009 & UBVR~(T06) & 0~  \\
SNLS-03D1au & 0.504 & griz~ & 0~  \\
SNLS-03D1aw & 0.582 & griz~ & 0~  \\
SNLS-03D1ax & 0.496 & griz~ & 1 ; \, $ p = -3 $~  \\
SNLS-03D1bk & 0.865 & griz~ & 1 ; \, $ p = -6 $~  \\
SNLS-03D1co & 0.679 & griz~ & 1 ; \, $ p = -4 $~  \\
SNLS-03D1ew & 0.868 & griz~ & 1 ; \, $ p = 1 $~  \\
SNLS-03D1fc & 0.331 & griz~ & 1 ; \, $ p = -3 $~  \\
SNLS-03D1fl & 0.688 & griz~ & 1 ; \, $ p = 1 $~  \\
SNLS-03D1fq & 0.800 & griz~ & 0~  \\
SNLS-03D4ag & 0.285 & griz~ & 1 ; \, $ p = -8 $~  \\
SNLS-03D4at & 0.633 & griz~ & 1 ; \, $ p = 6 $~  \\
SNLS-03D4cj & 0.270 & griz~ & 1 ; \, $ p = -9 $~  \\
SNLS-03D4cx & 0.949 & griz~ & 0~  \\
SNLS-03D4cy & 0.927 & griz~ & 0~  \\
SNLS-03D4cz & 0.695 & griz~ & 0~  \\
SNLS-03D4dh & 0.627 & griz~ & 0~  \\
SNLS-03D4dy & 0.604 & griz~ & 1 ; \, $ p = 4 $~  \\
SNLS-03D4fd & 0.791 & griz~ & 1 ; \, $ p = -1 $~  \\
SNLS-03D4gl & 0.571 & griz~ & 0~  \\
SNLS-04D1ag & 0.557 & griz~ & 1 ; \, $ p = 4 $~  \\
SNLS-04D1dc & 0.211 & griz~ & 1 ; \, $ p = -1 $~  \\
SNLS-04D1ff & 0.860 & griz~ & 1 ; \, $ p = 4 $~  \\
SNLS-04D1hd & 0.369 & griz~ & 1 ; \, $ p = 0 $~  \\
SNLS-04D1hx & 0.560 & griz~ & 1 ; \, $ p = 5 $~  \\
SNLS-04D1hy & 0.850 & griz~ & 1 ; \, $ p = -2 $~  \\
SNLS-04D1iv & 0.998 & griz~ & 1 ; \, $ p = 2 $~  \\
SNLS-04D1kj & 0.584 & griz~ & 1 ; \, $ p = -4 $~  \\
SNLS-04D1ks & 0.798 & griz~ & 1 ; \, $ p = -1 $~  \\
SNLS-04D1oh & 0.590 & griz~ & 0~  \\
SNLS-04D1ow & 0.915 & griz~ & 1 ; \, $ p = 5 $~  \\
SNLS-04D1pc & 0.758 & griz~ & 0~  \\
SNLS-04D1pd & 0.952 & griz~ & 0~  \\
SNLS-04D1pg & 0.515 & griz~ & 1 ; \, $ p = -1 $~  \\
SNLS-04D1pp & 0.735 & griz~ & 0~  \\
SNLS-04D1sa & 0.589 & griz~ & 1 ; \, $ p = -2 $~  \\
SNLS-04D1si & 0.704 & griz~ & 0~  \\
SNLS-04D1sk & 0.663 & griz~ & 0~  \\
SNLS-04D2ac & 0.348 & griz~ & 0~  \\
SNLS-04D2cf & 0.369 & griz~ & 1 ; \, $ p = 8 $~  \\
SNLS-04D2fp & 0.415 & griz~ & 1 ; \, $ p = 1 $~  \\
SNLS-04D2fs & 0.357 & griz~ & 1 ; \, $ p = 1 $~  \\
SNLS-04D2gb & 0.430 & griz~ & 0~  \\
SNLS-04D2gc & 0.521 & griz~ & 1 ; \, $ p = -4 $~  \\
SNLS-04D2gp & 0.707 & griz~ & 1 ; \, $ p = 2 $~  \\
SNLS-04D2kr & 0.744 & griz~ & 0~  \\
SNLS-04D3co & 0.620 & griz~ & 0~  \\
SNLS-04D3dd & 1.010 & griz~ & 1 ; \, $ p = 2 $~  \\
SNLS-04D3df & 0.470 & griz~ & 0~  \\
SNLS-04D3do & 0.610 & griz~ & 0~  \\
SNLS-04D3ez & 0.263 & griz~ & 0~  \\
SNLS-04D3fk & 0.358 & griz~ & 0~  \\
SNLS-04D3fq & 0.730 & griz~ & 1 ; \, $ p = 1 $~  \\
SNLS-04D3hn & 0.552 & griz~ & 0~  \\
SNLS-04D3kr & 0.337 & griz~ & 1 ; \, $ p = 4 $~  \\
SNLS-04D3mk & 0.813 & griz~ & 0~  \\
SNLS-04D3ml & 0.950 & griz~ & 0~  \\
SNLS-04D3nh & 0.340 & griz~ & 1 ; \, $ p = 3 $~  \\
SNLS-04D3ny & 0.810 & griz~ & 1 ; \, $ p = 1 $~  \\
SNLS-04D3oe & 0.756 & griz~ & 0~  \\
SNLS-04D4an & 0.613 & griz~ & 0~  \\
SNLS-04D4bq & 0.550 & griz~ & 2 ; \, $ p = 2,4 $~  \\
SNLS-04D4dm & 0.811 & griz~ & 0~  \\
SNLS-04D4fx & 0.629 & griz~ & 1 ; \, $ p = -8 $~  \\
SNLS-04D4gg & 0.424 & griz~ & 0~  \\
SNLS-04D4hu & 0.703 & griz~ & 0~  \\
SNLS-04D4ic & 0.680 & griz~ & 1 ; \, $ p = 2 $~  \\
SNLS-04D4ii & 0.866 & griz~ & 0~  \\
SNLS-04D4im & 0.751 & griz~ & 0~  \\
SNLS-04D4in & 0.516 & griz~ & 0~  \\
SNLS-04D4jr & 0.482 & griz~ & 1 ; \, $ p = -6 $~  \\
SNLS-05D1ck & 0.617 & griz~ & 0~  \\
SNLS-05D2ab & 0.320 & griz~ & 0~  \\
SNLS-05D2ac & 0.494 & griz~ & 0~  \\
SNLS-05D2ah & 0.180 & griz~ & 0~  \\
SNLS-05D2ay & 0.921 & griz~ & 0~  \\
SNLS-05D2bt & 0.679 & griz~ & 0~  \\
SNLS-05D2bv & 0.474 & griz~ & 0~  \\
SNLS-05D2bw & 0.921 & griz~ & 0~  \\
SNLS-05D2ci & 0.631 & griz~ & 0~  \\
SNLS-05D2cl & 0.839 & griz~ & 0~  \\
SNLS-05D2ct & 0.734 & griz~ & 0~  \\
SNLS-05D2dw & 0.417 & griz~ & 0~  \\
SNLS-05D2dy & 0.498 & griz~ & 0~  \\
SNLS-05D2eb & 0.639 & griz~ & 0~  \\
SNLS-05D2ec & 0.526 & griz~ & 0~  \\
SNLS-05D2fq & 0.735 & griz~ & 0~  \\
SNLS-05D2hc & 0.360 & griz~ & 0~  \\
SNLS-05D2he & 0.608 & griz~ & 0~  \\
SNLS-05D3ax & 0.643 & griz~ & 0~  \\
SNLS-05D3cf & 0.419 & griz~ & 0~  \\
SNLS-05D3ci & 0.515 & griz~ & 0~  \\
SNLS-05D3cx & 0.805 & griz~ & 0~  \\
SNLS-05D3dd & 0.480 & griz~ & 0~  \\
SNLS-05D3gp & 0.580 & griz~ & 0~  \\
SNLS-05D3gv & 0.715 & griz~ & 0~  \\
SNLS-05D3ha & 0.808 & griz~ & 0~  \\
SNLS-05D3hq & 0.338 & griz~ & 0~  \\
SNLS-05D3hs & 0.664 & griz~ & 0~  \\
SNLS-05D3ht & 0.900 & griz~ & 0~  \\
SNLS-05D3jb & 0.740 & griz~ & 0~  \\
SNLS-05D3jh & 0.718 & griz~ & 0~  \\
SNLS-05D3jk & 0.736 & griz~ & 0~  \\
SNLS-05D3jq & 0.579 & griz~ & 0~  \\
SNLS-05D3jr & 0.370 & griz~ & 0~  \\
SNLS-05D3kp & 0.850 & griz~ & 0~  \\
SNLS-05D3kt & 0.648 & griz~ & 0~  \\
SNLS-05D3kx & 0.219 & griz~ & 0~  \\
SNLS-05D3lb & 0.647 & griz~ & 1 ; \, $ p = -1 $~  \\
SNLS-05D3mq & 0.240 & griz~ & 0~  \\
SNLS-05D4af & 0.502 & griz~ & 0~  \\
SNLS-05D4ag & 0.636 & griz~ & 0~  \\
SNLS-05D4av & 0.543 & griz~ & 0~  \\
SNLS-05D4be & 0.540 & griz~ & 0~  \\
SNLS-05D4bj & 0.704 & griz~ & 0~  \\
SNLS-05D4ca & 0.608 & griz~ & 0~  \\
SNLS-05D4cq & 0.783 & griz~ & 0~  \\
SNLS-05D4cs & 0.735 & griz~ & 0~  \\
SNLS-05D4cw & 0.375 & griz~ & 0~  \\
SNLS-05D4dt & 0.407 & griz~ & 0~  \\
SNLS-05D4dw & 0.853 & griz~ & 0~  \\
SNLS-05D4dy & 0.810 & griz~ & 0~  \\
\enddata
\tablenotetext{a}{Heliocentric redshift.}
\tablenotetext{b}{Photometry References : 
B83:~\citet{Buta83},
P87:~\citet{Phillips87},
W94:~\citet{Wells94},
H96:~\citet{Hamuy96b},
L91:~\citet{Leibundgut91},
L98:~\citet{Lira98},
F92:~\citet{Filippenko92},
A04:~\citet{Altavilla04},
K04b:~\citet{Krisciunas04b},
S92:~\citet{Suntzeff92},
R99:~\citet{riess99a},
R95:~\citet{Richmond95},
P96:~\citet{Patat96},
M96:~\citet{Meikle96},
J05:~\citet{Jha05},
R05:~\citet{Riess05},
S99:~\citet{Suntzeff99},
St02:~\citet{Strolger02},
K00:~\citet{Krisciunas00},
K01:~\citet{Krisciunas01},
S02:~\citet{Stritzinger02},
V03:~\citet{Valentini03},
K04a:~\citet{Krisciunas04a},
K03:~\citet{Krisciunas03},
Vi03:~\citet{Vinko03},
Z03:~\citet{Zapata03},
B04:~\citet{Benetti04},
P04:~\citet{Pignata04},
A05:~\citet{Anupama05},
T06:~\citet{Tsvetkov06b}
}
\tablenotetext{c}{Only spectra with negligible host-contamination are listed. Spectroscopic References : 
Br83:~\citet{Branch83},
AC:~\citet{AsiagoCatalogue},
IUE:~\citet{IUE},
CFA:~\citet{CFA},
G04:~\citet{Garavini04},
H02:~\citet{Hamuy02},
W03:~\citet{Wang03sn01el},
B04:~\citet{Benetti04},
A05:~\citet{Anupama05}
}
\label{tab:training_sample}
\end{deluxetable}

\end{document}